\documentclass[12pt]{article}
\usepackage{graphicx}
\usepackage{lscape}
\usepackage{epsfig}
\usepackage{citesort}
\begin{document}
\def\ds{\displaystyle}
\def\beq{\begin{equation}}
\def\eeq{\end{equation}}
\def\bea{\begin{eqnarray}}
\def\eea{\end{eqnarray}}
\def\beeq{\begin{eqnarray}}
\def\eeeq{\end{eqnarray}}

\def\ve{\vert}
\def\vel{\left|}
\def\ver{\right|}

\def\nnb{\nonumber}
\def\ga{\left(}
\def\dr{\right)}
\def\aga{\left\{}
\def\adr{\right\}}
\def\lla{\left<}
\def\rra{\right>}
\def\rar{\rightarrow}
\def\nnb{\nonumber}
\def\la{\langle}
\def\ra{\rangle}
\def\ba{\begin{array}}
\def\ea{\end{array}}
\def\tr{\mbox{Tr}}
\def\ssp{{\Sigma^{*+}}}
\def\sso{{\Sigma^{*0}}}
\def\ssm{{\Sigma^{*-}}}
\def\xis0{{\Xi^{*0}}}
\def\xism{{\Xi^{*-}}}
\def\qs{\la \bar s s \ra}
\def\qu{\la \bar u u \ra}
\def\qd{\la \bar d d \ra}
\def\qq{\la \bar q q \ra}
\def\gGgG{\la g^2 G^2 \ra}
\def\q{\gamma_5 \not\!q}
\def\x{\gamma_5 \not\!x}
\def\g5{\gamma_5}
\def\sb{S_Q^{cf}}
\def\sd{S_d^{be}}
\def\su{S_u^{ad}}
\def\ss{S_s^{??}}
\def\sbp{{S}_Q^{'cf}}
\def\sdp{{S}_d^{'be}}
\def\sup{{S}_u^{'ad}}
\def\ssp{{S}_s^{'??}}
\def\sig{\sigma_{\mu \nu} \gamma_5 p^\mu q^\nu}
\def\fo{f_0(\frac{s_0}{M^2})}
\def\ffi{f_1(\frac{s_0}{M^2})}
\def\fii{f_2(\frac{s_0}{M^2})}
\def\O{{\cal O}}
\def\sl{{\Sigma^0 \Lambda}}
\def\es{\!\!\! &=& \!\!\!}
\def\ap{\!\!\! &\approx& \!\!\!}
\def\ar{&+& \!\!\!}
\def\ek{&-& \!\!\!}
\def\kek{\!\!\!&-& \!\!\!}
\def\cp{&\times& \!\!\!}
\def\se{\!\!\! &\simeq& \!\!\!}
\def\eqv{&\equiv& \!\!\!}
\def\kpm{&\pm& \!\!\!}
\def\kmp{&\mp& \!\!\!}
\def\md{\!\!\!\! &\mid& \!\!\!\!}

\def\lb{\Lambda_b}
\def\ll{\Lambda}
\def\mb{m_{\Lambda_b}}
\def\ml{m_\Lambda}
\def\s1{\hat s}
\def\ds{\displaystyle}

\def\OEE{\Omega_{\rm IB}}

\newcommand{\nnl}{\nonumber\\}
\newcommand{\kk}[1]{_{(#1)}}  
\newcommand{\kkx}[2]{_{#1(#2)}}  
\newcommand{\parity}{\mathbf{P}}
\newcommand{\mcD}{\mathcal{D}}
\newcommand{\mcE}{\mathcal{E}}
\newcommand{\mcG}{\mathcal{G}}
\newcommand{\mcL}{\mathcal{L}}
\newcommand{\mcQ}{\mathcal{Q}}
\newcommand{\mcU}{\mathcal{U}}
\newcommand{\fslash}[1]{#1\!\!\!/}  
\newcommand{\nor}{\frac{n}{R}} 
\newcommand{\norsq}{\frac{n^2}{R^2}} 
\newcommand{\GeV}{~\text{GeV}}
\newcommand{\dpkt}{\;:\quad}
\newcommand{\sw}{s_w}  
\newcommand{\cw}{c_w}

\newcommand{\RE}{{\rm Re}}
\newcommand{\IM}{{\rm Im}}
\newcommand{\vcb}{|V_{cb}|}
\newcommand{\vtd}{|V_{td}|}
\newcommand{\vub}{|V_{ub}/V_{cb}|}
\newcommand{\vts}{|V_{ts}|}
\newcommand{\vus}{|V_{us}|}

\newcommand{\tvs}{\vbox{\vskip 3mm}}
\newcommand{\svs}{\vbox{\vskip 5mm}}
\newcommand{\mvs}{\vbox{\vskip 8mm}}
\newcommand{\msvs}{\vbox{\vskip 7mm}}

\def\ss{\langle \bar s s \rangle}
\def\qq{\langle \bar q q \rangle}
\def\mixedss{\langle \bar s \sigma g_s G s \rangle}
\newcommand{\nn}{\nonumber}
\newcommand{\dd}{\displaystyle}
\newcommand{\bra}[1]{\left\langle #1 \right|}
\newcommand{\ket}[1]{\left| #1 \right\rangle}
\newcommand{\spur}[1]{\not\! #1 \,}
\def\R1{\varepsilon_1}
\def\E8{\varepsilon_8}
\def\gat{\tilde{\gamma}}
\def\gh{\hat{g}}
\def\gt{\tilde{g}}
\def\gah{\hat{\gamma}}
\def\ga{\gamma}
\def\gaf{\gamma_{5}}

\def\lb{\Lambda_b}
\def\ll{\Lambda}
\def\mb{m_{\Lambda_b}}
\def\ml{m_\Lambda}
\def\s1{\hat s}
\def\ds{\displaystyle}

\def\eps{\varepsilon}
\def\epe{\varepsilon'/\varepsilon}
\def\as{\alpha_s}
\newcommand{\eqn}{\ref}
\def\Heff{{\cal H}_{\rm eff}}
\newcommand{\mt}{m_{\rm t}}
\newcommand{\mtb}{\overline{m}_{\rm t}}
\newcommand{\mcb}{\overline{m}_{\rm c}}
\newcommand{\mc}{m_{\rm c}}
\newcommand{\ms}{m_{\rm s}}
\newcommand{\mn}{m_{\rm n}}
\newcommand{\mw}{M_{\rm W}}
\newcommand{\mz}{M_{\rm Z}}

\newcommand{\gev}{\, {\rm GeV}}
\newcommand{\mev}{\, {\rm MeV}}
\newcommand{\bsi}{B_6^{(1/2)}}
\newcommand{\bei}{B_8^{(3/2)}}
\newcommand{\Lms}{\Lambda_{\overline{\rm MS}}}
\newcommand{\bsg}{$b \to s \gamma$ }
\newcommand{\Bsg}{$B \to X_s \gamma$ }
\newcommand{\newsection}[1]{\section{#1}\setcounter{equation}{0}}
\newcommand{\bd}{\begin{displaymath}}
\newcommand{\ed}{\end{displaymath}}
\newcommand{\aem}{\alpha}
\newcommand{\Bsee}{$B \to X_s e^+ e^-$ }
\newcommand{\bsee}{$b \to s e^+ e^-$ }
\newcommand{\bcenu}{$b \to c e \bar\nu $ }

\newcommand{\ord}{{\cal O}}
\newcommand{\order}{{\cal O}}
\newcommand{\f}{\frac}
\newcommand{\Ctilde}{\tilde{C}}
\newcommand{\kpnn}{K^+\rightarrow\pi^+\nu\bar\nu}
\newcommand{\kpn}{K^+\rightarrow\pi^+\nu\bar\nu}
\newcommand{\klpn}{K_{\rm L}\rightarrow\pi^0\nu\bar\nu}
\newcommand{\klpnn}{K_{\rm L}\rightarrow\pi^0\nu\bar\nu}
\newcommand{\klm}{K_{\rm L} \to \mu^+\mu^-}
\newcommand{\kmm}{K_{\rm L} \to \mu^+ \mu^-}
\newcommand{\kpe}{K_{\rm L} \to \pi^0 e^+ e^-}

\setlength{\unitlength}{1mm}
\textwidth 16.3 true cm
\textheight 23.0 true cm
\topmargin -0.8 true in
\oddsidemargin 0.00 true in
\def\theequation{\thesection.\arabic{equation}}
\def\OEE{\Omega_{\rm IB}}

\def\R1{\varepsilon_1}
\def\E8{\varepsilon_8}
\def\gat{\tilde{\gamma}}
\def\gh{\hat{g}}
\def\gt{\tilde{g}}
\def\gah{\hat{\gamma}}
\def\ga{\gamma}
\def\gaf{\gamma_{5}}
\def\eps{\varepsilon}
\def\epe{\varepsilon'/\varepsilon}
\def\as{\alpha_s}
\def\Heff{{\cal H}_{\rm eff}}

\newcommand{\imlt}{\IM\lambda_t}
\newcommand{\relt}{\RE\lambda_t}
\newcommand{\relc}{\RE\lambda_c}
\renewcommand{\baselinestretch}{1.3}
\renewcommand{\thefootnote}{\fnsymbol{}}
\def\ds{\displaystyle}
\def\beq{\begin{equation}}
\def\eeq{\end{equation}}
\def\bea{\begin{eqnarray}}
\def\eea{\end{eqnarray}}
\def\beeq{\begin{eqnarray}}
\def\eeeq{\end{eqnarray}}
\def\ve{\vert}
\def\vel{\left|}
\def\ver{\right|}
\def\nnb{\nonumber}
\def\ga{\left(}
\def\dr{\right)}
\def\aga{\left\{}
\def\adr{\right\}}
\def\lla{\left<}
\def\rra{\right>}
\def\rar{\rightarrow}
\def\nnb{\nonumber}
\def\la{\langle}
\def\ra{\rangle}
\def\ba{\begin{array}}
\def\ea{\end{array}}
\def\tr{\mbox{Tr}}
\def\ssp{{\Sigma^{*+}}}
\def\sso{{\Sigma^{*0}}}
\def\ssm{{\Sigma^{*-}}}
\def\xis0{{\Xi^{*0}}}
\def\xism{{\Xi^{*-}}}
\def\qs{\la \bar s s \ra}
\def\qu{\la \bar u u \ra}
\def\qd{\la \bar d d \ra}
\def\qq{\la \bar q q \ra}
\def\gGgG{\la g^2 G^2 \ra}
\def\q{\gamma_5 \not\!q}
\def\x{\gamma_5 \not\!x}
\def\g5{\gamma_5}
\def\sb{S_Q^{cf}}
\def\sd{S_d^{be}}
\def\su{S_u^{ad}}
\def\ss{S_s^{??}}
\def\sbp{{S}_Q^{'cf}}
\def\sdp{{S}_d^{'be}}
\def\sup{{S}_u^{'ad}}
\def\ssp{{S}_s^{'??}}
\def\sig{\sigma_{\mu \nu} \gamma_5 p^\mu q^\nu}
\def\fo{f_0(\frac{s_0}{M^2})}
\def\ffi{f_1(\frac{s_0}{M^2})}
\def\fii{f_2(\frac{s_0}{M^2})}
\def\O{{\cal O}}
\def\sl{{\Sigma^0 \Lambda}}
\def\es{\!\!\! &=& \!\!\!}
\def\ap{\!\!\! &\approx& \!\!\!}
\def\ar{&+& \!\!\!}
\def\ek{&-& \!\!\!}
\def\kek{\!\!\!&-& \!\!\!}
\def\cp{&\times& \!\!\!}
\def\se{\!\!\! &\simeq& \!\!\!}
\def\eqv{&\equiv& \!\!\!}
\def\kpm{&\pm& \!\!\!}
\def\kmp{&\mp& \!\!\!}
\def\simlt{\stackrel{<}{{}_\sim}}
\def\simgt{\stackrel{>}{{}_\sim}}

\newcommand{\me}[1]{\langle#1\rangle}
\newcommand{\al}{\alpha_s}

\def\simlt{\stackrel{<}{{}_\sim}}
\def\simgt{\stackrel{>}{{}_\sim}}


\title{
         {\Large
                 {\bf Semileptonic Transition
 of $\Sigma_b \rar \Sigma \mu^+ \mu^-$ in Family
Non-universal $Z^\prime$ Model
                 }
         }
      }

\author{\vspace{1cm}\\
{ N. Kat{\i}rc{\i}$^1$\thanks
{e-mail: nihan.katirci@boun.edu.tr}} \,, K. Azizi$^2$ \thanks {e-mail: kazizi@dogus.edu.tr}  \\
{\small $^1$ Department of Physics, Bo\u gazi\c ci University, 34470 Bebek, Istanbul, Turkey }\\
{\small $^2$ Department of Physics, Do\u gu\c s
University}, {\small Ac{\i}badem-Kad{\i}k\"oy, 34722 Istanbul, Turkey }\\
       }
\date{}
\begin{titlepage}
\maketitle \thispagestyle{empty}

\begin{abstract}
Using newly available form factors obtained from light cone QCD
sum rules in full theory, we study the flavor changing neutral
current transition of $\Sigma_b \rar \Sigma \mu^+ \mu^-$ decay in
the family non-universal $Z^\prime$ model. In particular, we
evaluate the differential branching ratio, forward-backward
asymmetry as well as some related asymmetry parameters and
polarizations. We compare the obtained results with the
predictions of the standard model and discuss the sensitivity of the observables under consideration
to family non-universal $Z^\prime$ gauge boson. The order of differential branching ratio
shows that this decay mode can be checked at LHC in near future.
\end{abstract}
~~~PACS numbers: 12.60.-i, 12.60.Cn, 13.30.-a, 13.30.Ce, 14.20.Mr
\end{titlepage}

\section{Introduction}

The heavy baryons containing a single heavy quark constitute a
perfect laboratory to test the non-perturbative aspects of QCD.
The $\Lambda_b \rar \Lambda \mu^+ \mu^-$ transition which is
based on the flavor changing neutral current (FCNC) transition of
$b \rightarrow s\mu^+\mu^-$ at quark level has been recently observed
in CDF collaboration at FermiLab \cite{cdfobs}. It is also planned
to be checked at LHCb collaboration at CERN \cite{lhcb}.

The theoretical studies on the branching ratio of $\Sigma_b \rar
\Sigma \ell^+ \ell^-$ in standard model (SM) \cite{formfactor}
show that this decay mode is also possible to be observed at LHC.
It is expected that the experimental studies on the heavy baryons
and their decay properties constitute one of the main direction of
research program at LHC. Hence, the theoretical calculations can
play an essential role in this regard.

Although the SM has predictions in perfect agreement with collider
data up to now, there are some problems such as neutrino
oscillations, baryon asymmetry, unification, dark matter, strong
CP violation and the hierarchy problem, etc. which can not be
addressed by the SM and still remain unsolved. To cure these
deficiencies, there are a plenty of new physics (NP) models such
as different extra dimension models (ED), various supersymmetric
(SUSY) scenarios, etc. One of the most important new physics
scenarios is $Z^\prime$ model, appears in many grand unified
theories, such as $SU(5)$ or string-inspired $E6$ models
\cite{ealti,buchalla,nardi,bernabeu,bargerv}. The two $Z^\prime$
models in agenda are family non-universal $Z^\prime$
\cite{theoz,zp} and leptophobic $Z^\prime$ scenarios
\cite{lepto,sirvanli}.

\bigskip The idea of extra heavy Z boson comes from the extension
of gauge group $SU(5)$, predicted by the grand unification
theories to larger group $SO(10)$. The $SO(10)$ gauge group is the
next important one after $SU(5)$ having one extra rank. Hence,
this gauge group requires at least one extra neutral gauge boson
\cite{leike}. In general, $Z^\prime$ gauge couplings are family
universal
\cite{zp,nonunibir,nonuniiki,nonuniuc,nonunid,nonunib,nonunia},
however, due to different constructions of the different families,
in string models it is possible to have family non-universal
$Z^\prime$ couplings. In some of them, three generation of leptons
and also the first and second generation of quarks have different
coupling to $Z^\prime$ boson when compared to the third families
of quarks \cite{zp,nonu,nonuiki}. For more information about this
model see for instance \cite{theoz,zp,stu,ew,erler,fitb,paul}. The
leptophobic $Z^\prime$ model implies that the new neutral gauge
boson does not couple to the ordinary SM charged leptons.

 The study of the $Z^\prime$ phenomenology is
an important part of the scientific program of every present and
future colliders and due to its heaviness, the $Z^\prime$ boson may be used
to calibrate the future detectors \cite{leike}. For constraints on the
mass of the $Z^\prime$ boson and the mixing parameters of the model see
for example \cite{const1,constiki,cdf}. There are direct searches
for $Z^\prime \rightarrow e^+e^-$ decay \cite{directsearch} at
Tevatron, and the possibility to discover this gauge boson
is analyzed in \cite{tevatron}.

In the present work we investigate the FCNC transition of
$\Sigma_b \rar \Sigma \mu^+ \mu^-$ in family non-universal
$Z^\prime$ model. In particular, we analyze the differential
branching ratio, forward-backward asymmetry as well as some
related asymmetry parameters and polarizations and compare the results with
the predictions of the SM. Note that the rare baryonic $\Lambda_b
\rar \Lambda \ell^+\ell^-$ decay within family non-universal
$Z^\prime$ model was analyzed in
\cite{giri,alievzprime,alievzppol} within also family
non-universal $Z^\prime$ model. The implications of non-universal
$Z^\prime$ model on B meson decays were investigated in
\cite{zpB2,barger,zpB3,che,mohanta,hua,zpB4,li,wang}. The $B
\rightarrow K_2( \rightarrow K \pi)l^+ l^-$  \cite{bk2}, $B
\rightarrow K_1 l^+l^−$, and $B \rightarrow K_0 \pi$
\cite{bk1,bk0} were investigated in the same framework as well.
The effects of a family non-universal $Z^\prime$ gauge boson is
also searched for $B \rightarrow \pi\pi$ decays in \cite{btopi}.
Recently, the $B^0_s-\bar B^0_s$ mixing, $B \rightarrow K^* l^+l^-$, $B_s \rightarrow \mu^+\mu^-$, $B \rightarrow K \pi$ and inclusive $B \rightarrow X_sl^+l^-$ 
decays have been analyzed and the stronger constraints have been put for the family non-universal $Z^\prime$ model parameters in \cite{obserbmixing,abazov,aaltonen,aajione,aajisec} 
(see also \cite{alievzprime,alievzppol}).

The outline of the paper is as follows. In section 2, we present
the effective Hamiltonian and transition matrix elements
responsible for the $\Sigma_b \rar \Sigma \mu^+ \mu^-$ transition.
In section 3, we analyze the differential branching ratio, forward
backward asymmetry, double lepton polarizations as well as some
other related asymmetries in the $Z^\prime$ model and compare the
obtained results with the SM predictions. The last section
encompasses our concluding remarks.

\section{The $\Sigma_b \rar \Sigma \mu^+ \mu^-$
transition in Family Non-universal $Z^\prime$ Model}
\subsection{ The Effective Hamiltonian}
Neglecting terms proportional to $\frac{V_{ub} V_{us}^*} {V_{tb}
V_{ts}^*} \approx O(10^{-2})$, the effective Hamiltonian of the
FCNC transition of $\Sigma_b \rar \Sigma \ell^+ \ell^-$, proceed
via quark level $b \rar s \ell^+ \ell^-$ in the SM, can be written
as \cite{breitwigner,bobeth,altmann,ghin}
 \bea \label{e8401} {\cal H}^{eff} &=& {G_F \alpha_{em} V_{tb}
V_{ts}^\ast \over 2\sqrt{2} \pi} \Bigg[ C_9^{eff}
\bar{s}\gamma_\mu (1-\gamma_5) b \, \bar{\ell} \gamma^\mu \ell +
C_{10}  \bar{s} \gamma_\mu (1-\gamma_5) b \, \bar{\ell}
\gamma^\mu
\gamma_5 \ell \nnb \\
&-&  2 m_b C_7^{eff}  {1\over q^2} \bar{s} i
\sigma_{\mu\nu}q^{\nu} (1+\gamma_5) b \, \bar{\ell} \gamma^\mu
\ell \Bigg]~, \eea where $\alpha_{em}$  is the fine structure
constant at Z mass scale, $G_F$ is the Fermi coupling constant,
$V_{ij}$ are elements of the Cabibbo-Kobayashi-Maskawa (CKM)
matrix, and $C_7^{eff}$, $C_9^{eff}$, $C_{10}$ are the Wilson
coefficients. When $Z^\prime$ boson is considered and $Z-Z^\prime$
mixing is neglected, the extra part which should be added to the
above effective Hamiltonian is written as  \cite{zpB2,barger}
 \bea \label{e111203}
H_{eff}^{Z^\prime} \es- {2 G_F \over \sqrt{2}} V_{tb} V_{ts}^\ast
\Bigg[ {B_{sb}^L B_{\ell\ell}^L \over V_{tb} V_{ts}^\ast } \bar{s}
\gamma_\mu (1-\gamma_5) b \, \bar{\ell} \gamma_\mu (1-\gamma_5)
\ell \nnb\\ \ar {B_{sb}^L B_{\ell\ell}^R \over V_{tb} V_{ts}^\ast } \bar{s}
\gamma_\mu (1-\gamma_5) b \, \bar{\ell} \gamma_\mu (1+\gamma_5)
\ell \bigg]+h.c.~ \nnb \\
 \eea where $B_{sb}^L= |B_{sb}^L|e^{i\varphi_s^L}$ and $B_{\ell\ell}^{L,R}$
correspond to the chiral $Z^\prime$ couplings to quarks and
leptons, respectively. Considering the running effects from $m_W$ to $m_b$ scale \cite{misiak}, 
 to get the
effective Hamiltonian for the transition under consideration in $Z^\prime$ model, we
need to make the following replacements in Eq. (\ref{e8401}) to include
$Z^\prime$ boson contributions besides the Z boson: \bea
\label{e111207} C_9^{\rm eff} &\rar& C_9^{\rm eff \prime}=C_9^{\rm
eff} - \frac{4\pi}{\alpha_s}(28.82) {B_{sb}^L \over V_{tb}
V_{ts}^\ast }(B_{\ell\ell}^L +
B_{\ell\ell}^R), \nnb \\
C_{10} &\rar& C_{10}^{\prime}=C_{10} +
\frac{4\pi}{\alpha_s}(28.82) {B_{sb}^L \over V_{tb} V_{ts}^\ast }
(B_{\ell\ell}^L - B_{\ell\ell}^R), \eea where $\alpha_s$ is the
strong coupling constant. Here we should mention that the Wilson
coefficient $C_7^{eff}$ remains unchanged.

 In SM, the Wilson coefficient $C_7^{eff}$ in leading logarithm approximation is written as (see \cite{cyedi})
  \bea
\label{cyedi} C_7^{eff}(\mu_b) \es \eta^{\frac{16}{23}}
C_7(\mu_W)+ \frac{8}{3} \left( \eta^{\frac{14}{23}}
-\eta^{\frac{16}{23}} \right) C_8(\mu_W)+C_2 (\mu_W) \sum_{i=1}^8
h_i \eta^{a_i}~, \eea where \bea
 C_2(\mu_W)=1~,~~ C_7(\mu_W)=-\frac{1}{2}
D_0(x_t)~,~~ C_8(\mu_W)=-\frac{1}{2} E_0(x_t)~ . \eea
 The functions, $D_0(x_t)$  and $E_0(x_t)$ are defined as
 \bea \label{d0} D_0(x_t) \es - \frac{(8
x_t^3+5 x_t^2-7 x_t)}{12 (1-x_t)^3}
+ \frac{x_t^2(2-3 x_t)}{2(1-x_t)^4}\ln x_t~, \\ \nnb \\
\label{e0} E_0(x_t) \es - \frac{x_t(x_t^2-5 x_t-2)}{4 (1-x_t)^3} +
\frac{3 x_t^2}{2 (1-x_t)^4}\ln x_t~, \eea where
$x_{t}=\frac{m_{t}^{2}}{M_{W}^{2}}$ with $m_t$ and $M_W$ being the
top quark and W boson masses, respectively.
 The coefficients $a_i$
and $h_i$ are given as  \cite{cdokuz,cdok} \bea \frac{}{}
   \label{klar}
\begin{array}{rrrrrrrrrl}
a_i = (\!\! & \f{14}{23}, & \f{16}{23}, & \f{6}{23}, & -
\f{12}{23}, &
0.4086, & -0.4230, & -0.8994, & 0.1456 & \!\!)  \vspace{0.1cm}, \\
h_i = (\!\! & 2.2996, & - 1.0880, & - \f{3}{7}, & - \f{1}{14}, &
-0.6494, & -0.0380, & -0.0186, & -0.0057 & \!\!).
\end{array}
\eea The parameter $\eta$ in Eq. (\ref{cyedi}) is also defined as
 \bea \eta \es
\frac{\alpha_s(\mu_W)} {\alpha_s(\mu_b)}~,\eea
 with
 \bea \alpha_s(x)=\frac{\alpha_s(m_Z)}{1-\beta_0\frac{\alpha_s(m_Z)}{2\pi}\ln(\frac{m_Z}{x})},\eea
where $\alpha_s(m_Z)=0.118$ and $\beta_0=\frac{23}{3}$.

The Wilson coefficient $C_{9}^{eff}(\hat s')$ is written as
\cite{cdokuz,cdok}:
 \bea \label{C9eff}
C_{9}^{eff}(\hat{s}') & = & C_9^{NDR}\eta(\hat s') + h(z, \hat s')\left( 3
C_1 + C_2 + 3 C_3 + C_4 + 3
C_5 + C_6 \right) \nonumber \\
& & - \f{1}{2} h(1, \hat s') \left( 4 C_3 + 4 C_4 + 3
C_5 + C_6 \right) \nonumber \\
& & - \f{1}{2} h(0, \hat s') \left( C_3 + 3 C_4 \right)
+ \f{2}{9} \left( 3 C_3 + C_4 + 3 C_5 +
C_6 \right), \eea

where $\hat{s}'=\frac{q^2}{m_b^2}$ with $4m_l^2\leq
q^2\leq(m_{\Sigma_b}-m_\Sigma)^2$, and \bea
\label{C9tilde}C_9^{NDR} & = & P_0^{NDR} +
\f{Y_0(x_t)}{\sin^2\theta_W} -4 Z_0(x_t) + P_E E(x_t), \eea
 here, NDR stands for the naive dimensional regularization scheme. We ignore the last term in this equation
 due to the negligible value of $P_E$. The  $P_0^{NDR}=2.60 \pm 0.25$
\cite{cdokuz,cdok} and the functions $Y_0(x_t)$ and $Z_0(x_t)$ are
defined in the following form:
 \bea \label{yo} Y_0(x_t) \es \frac{x_t}{8} \left[ \frac{x_t
-4}{x_t -1}+\frac{3 x_t}{(x_t-1)^2} \ln x_t \right],\eea and \bea
Z_0(x_t) \es \frac{18 x_t^4-163 x_t^3+259 x_t^2 -108 x_t}{144
(x_t-1)^3}+\left[\frac{32 x_t^4-38 x_t^3-15 x_t^2+18 x_t}{72
(x_t-1)^4} - \frac{1}{9}\right] \ln x_t. \nonumber \\\eea

 In Eq.(\ref{C9eff}), the $\eta(\hat s')$ is given as
 \bea \eta(\hat s') & = & 1 + \f{\al(\mu_b)}{\pi}\,
\omega(\hat s'), \eea with
 \bea \label{omega}
\omega(\hat s') & = & - \f{2}{9} \pi^2 - \f{4}{3}\mbox{Li}_2(\hat s') -
\f{2}{3}
\ln \hat s' \ln(1-\hat s') - \f{5+4\hat s'}{3(1+2\hat s')}\ln(1-\hat s') - \nonumber \\
& &  \f{2 \hat s' (1+\hat s') (1-2\hat s')}{3(1-\hat s')^2
(1+2\hat s')} \ln \hat s' + \f{5+9\hat s'-6\hat s'^2}{6 (1-\hat
s') (1+2\hat s')}. \eea At $\mu_b$ scale, for the coefficients
$C_j~(j=1,...6)$ we have \bea \label{coeffs} C_j=\sum_{i=1}^8
k_{ji} \eta^{a_i}, \vspace{0.2cm} \eea where $k_{ji}$ are given
as: \bea \frac{}{}
   \label{klar}
\begin{array}{rrrrrrrrrl}
k_{1i} = (\!\! & 0, & 0, & \f{1}{2}, & - \f{1}{2}, &
0, & 0, & 0, & 0 & \!\!),  \vspace{0.1cm} \\
k_{2i} = (\!\! & 0, & 0, & \f{1}{2}, &  \f{1}{2}, &
0, & 0, & 0, & 0 & \!\!),  \vspace{0.1cm} \\
k_{3i} = (\!\! & 0, & 0, & - \f{1}{14}, &  \f{1}{6}, &
0.0510, & - 0.1403, & - 0.0113, & 0.0054 & \!\!),  \vspace{0.1cm} \\
k_{4i} = (\!\! & 0, & 0, & - \f{1}{14}, &  - \f{1}{6}, &
0.0984, & 0.1214, & 0.0156, & 0.0026 & \!\!),  \vspace{0.1cm} \\
k_{5i} = (\!\! & 0, & 0, & 0, &  0, &
- 0.0397, & 0.0117, & - 0.0025, & 0.0304 & \!\!) , \vspace{0.1cm} \\
k_{6i} = (\!\! & 0, & 0, & 0, &  0, &
0.0335, & 0.0239, & - 0.0462, & -0.0112 & \!\!).  \vspace{0.1cm} \\
\end{array}
\eea The other functions in Eq. (\ref{C9eff}) are also given as:
 \bea \label{phasespace} h(y,
\hat s') & = & -\f{8}{9}\ln\f{m_b}{\mu_b} - \f{8}{9}\ln y +
\f{8}{27} + \f{4}{9} x \nonumber \\
& & - \f{2}{9} (2+x) |1-x|^{1/2} \left\{
\begin{array}{ll}
\left( \ln\left| \f{\sqrt{1-x} + 1}{\sqrt{1-x} - 1}\right| - i\pi
\right), &
\mbox{for } x \equiv \f{4z^2}{\hat s'} < 1 \nonumber \\
2 \arctan \f{1}{\sqrt{x-1}}, & \mbox{for } x \equiv \f {4z^2}{\hat
s'} > 1,
\end{array}
\right. \\
\eea
where   $y=1$ or $y=z=\frac{m_c}{m_b}$ and,
\bea h(0, \hat s') & = & \f{8}{27}
-\f{8}{9} \ln\f{m_b}{\mu_b} - \f{4}{9} \ln \hat s' + \f{4}{9}
i\pi.\eea
We also consider the long distance contribution ($Y_{LD}$) coming
from $J/\psi$ family resonances to $C_9^{eff}$ and parameterized
using Breit-Wigner ansatz \cite{breitwigner} as \bea
Y_{LD}=\frac{3\pi}{\alpha_{em}^2}C^{(0)}{\sum_{i=1}^6}\kappa_i
\frac{\Gamma (V_i \rightarrow l^+ l^-)m_{V_i}}{m_{V_i}^2-q^2-i
m_{V_i} \Gamma_{V_i}}, \eea where $C^{(0)}=0.362$, and we consider
only two lowest resonances $J/\psi (1S)$ and $\psi (2S)$ and
choose the corresponding phenomenological factors $\kappa_1=1$ and
$\kappa_2=2$. We use the experimental results on the masses and
total decay rates of dilepton decays of the considered vector
charmonium states \cite{char}. For more details about the
calculation of long distance contributions, see for instance
\cite{longbir,longiki}.

Finally, the explicit expression for $C_{10}$ is given as:
 \bea
\label{con} C_{10}= - \frac{Y_0(x_t)}{\sin^2 \theta_W}~, \eea
where, $\sin^2\theta_W= 0.23$.
\subsection{Transition Matrix Elements and Form Factors}
 The  $\Sigma_b \rar \Sigma \mu^+ \mu^-$ decay amplitude is obtained by sandwiching
the effective Hamiltonian between the initial and final states
\begin{eqnarray} \label{amplitude}
 {\cal M}  \es  \langle \Sigma(p) \vert H_{\rm eff} \vert \Sigma_b (p+q) \rangle,
\end{eqnarray} which reads
\begin{eqnarray}
\label{amplitude}
{\cal M} = \frac{G_F~\alpha_{em}
V_{tb}~V_{ts}^{^{*}}}{2\sqrt2~\pi} \Bigg\{\vphantom{\int_0^{x_2}}
{C_{9}^{\rm eff}}^{\prime}~ \langle \Sigma(p) \vert \bar{s}
\gamma_\mu (1-\gamma_5) b \vert \Sigma_b(p+q) \rangle \bar l \gamma^\mu l \nnb \\
+ C_{10}^{\prime} ~\langle \Sigma(p) \vert \bar{s}
\gamma_\mu (1-\gamma_5) b \vert \Sigma_b(p+q) \rangle \bar l
\gamma^\mu \gamma_{5}l  \nnb \\
-2 m_{b}~C_{7}^{\rm eff}\frac{1}{q^{2}}
~\langle \Sigma(p) \vert \bar{s} i \sigma_{\mu\nu}q^{\nu}
(1+\gamma_5) b \vert \Sigma_b(p+q) \rangle \bar l \gamma^\mu l
\Bigg\}~.
\end{eqnarray}
To calculate the amplitude, we need to parameterize the transition
matrix elements  $\langle \Sigma(p) \vert \bar{s} \gamma_\mu
(1-\gamma_5) b \vert \Sigma_b(p+q) \rangle$ and $\langle \Sigma(p)
\vert \bar{s} \sigma_{\mu \nu} q^\nu (1+\gamma_5) b \vert
\Sigma_b(p+q) \rangle$ in terms of twelve form factors $f_i$,
$g_i$, $f^T_i$ and $g^T_i$ ($i=1,2,3$) as follows:
\begin{eqnarray}\label{mat1}
\langle \Sigma(p) \md  \bar s \gamma_\mu
(1-\gamma_5) b \mid \Sigma_b(p+q)\rangle= \bar {u}_\Sigma(p)
\Big[\gamma_{\mu}f_{1}(q^{2})+{i}
\sigma_{\mu\nu}q^{\nu}f_{2}(q^{2}) + q^{\mu}f_{3}(q^{2}) \nnb \\
\ek \gamma_{\mu}\gamma_5
g_{1}(q^{2})-{i}\sigma_{\mu\nu}\gamma_5q^{\nu}g_{2}(q^{2}) -
q^{\mu}\gamma_5 g_{3}(q^{2}) \vphantom{\int_0^{x_2}}\Big]
u_{\Sigma_{b}}(p+q)~, \nnb \\
\langle \Sigma(p)\md \bar s i
\sigma_{\mu\nu}q^{\nu} (1+ \gamma_5) b \mid \Sigma_b(p+q)\rangle
=\bar{u}_\Sigma(p)
\Big[\gamma_{\mu}f_{1}^{T}(q^{2})+{i}\sigma_{\mu\nu}q^{\nu}f_{2}^{T}(q^{2})+
q^{\mu}f_{3}^{T}(q^{2}) \nnb \\
\ar \gamma_{\mu}\gamma_5
g_{1}^{T}(q^{2})+{i}\sigma_{\mu\nu}\gamma_5q^{\nu}g_{2}^{T}(q^{2}) +
q^{\mu}\gamma_5 g_{3}^{T}(q^{2}) \vphantom{\int_0^{x_2}}\Big]
u_{\Sigma_{b}}(p+q)~,
\end{eqnarray}
where  $u_{\Sigma_b}$ and $u_{\Sigma}$ are spinors of $\Sigma_b$
and $\Sigma$ baryons, respectively.

The form factors as the main inputs in the analysis of the
$\Sigma_{b}\rightarrow \Sigma \ell^{+}\ell^{-}$ have been 
recently calculated in full theory via light cone QCD sum rules in
\cite{formfactor} (for details about the light cone QCD sum rules see for instance \cite{bir,iki,uc}).
 The fit function of transition form factors is given as
\begin{eqnarray}
f^{(T)}_i(q^2)[g^{(T)}_i(q^2)]=\frac{a}{(1-\frac{q^2}{m_{fit}^2})}+\frac{b}{(1-\frac{q^2}{m_{fit}^2})^2},
\label{parametrization1}
\end{eqnarray}
where the fit parameters $a$, $b$, and $m_{fit}$ are presented in
Table 1.
\begin{table}[h!]
\renewcommand{\arraystretch}{1.5}
\addtolength{\arraycolsep}{3pt}
$$
\begin{array}{|c|c|c|c|c|c|}

\hline \hline
                & \mbox{a} & \mbox{b}  & m_{fit}& q^2=0
                \\
\hline
 f_1            &  -(0.035\pm0.006) &   0.130\pm0.023   &  5.1\pm1.0   &  0.095 \pm 0.017  \\
 f_2            &  0.026\pm0.006  &  -(0.081\pm0.018)  &  5.2 \pm1.0  & -0.055 \pm 0.012  \\
 f_3            &   0.013\pm0.004 &  -(0.065\pm0.020)  &  5.3 \pm1.1  & -0.052 \pm 0.016  \\
 g_1            &  -(0.031\pm0.008) &   0.151\pm0.038   &  5.3 \pm1.1  &  0.121 \pm 0.031    \\
 g_2            &  0.015\pm0.005  &  -(0.040\pm0.013)  &  5.3 \pm1.1  & -0.025 \pm 0.008  \\
 g_3            &  0.012\pm0.003  &  -(0.047\pm0.012)  &  5.4 \pm1.1  & -0.035 \pm 0.009  \\
 f_1^{T}        &  1.0\pm0.0    &   -(1.0\pm0.0)   &  5.4 \pm1.1  &  0.0\pm0.0        \\
 f_2^{T}        &  -(0.290\pm0.089)  &   0.421\pm0.129   &  5.4 \pm1.1  &  0.131 \pm 0.041    \\
 f_3^{T}        &  -(0.240\pm0.071)  &   0.412\pm0.122   &  5.4 \pm1.1  &  0.172 \pm 0.051    \\
 g_1^{T}        &  0.450\pm0.135   &   -(0.460\pm 0.138)  &  5.4  \pm1.1 & -0.010\pm 0.003   \\
 g_2^{T}        &  0.031\pm0.009  &   0.055\pm0.015  &  5.4  \pm1.1 &  0.086 \pm 0.024  \\
 g_3^{T}        &  -(0.011\pm0.003) &  -(0.180\pm0.057)   &  5.4  \pm1.1 & -0.190 \pm 0.060    \\
\hline \hline
\end{array}
$$
\caption{Parameters appearing in  the fit function of the  form
factors, $f_{1}$, $f_{2}$, $f_{3}$, $g_{1}$, $g_{2}$, $g_{3}$,
$f^T_{1}$, $f^T_{2}$, $f^T_{3}$, $g^T_{1}$, $g^T_{2}$ and
$g^T_{3}$ in full theory for $\Sigma_{b}\rightarrow
\Sigma\ell^{+}\ell^{-}$ together with the
  values of the form factors at $q^2=0$ \cite{formfactor}.} \label{tab:13}
\renewcommand{\arraystretch}{1}
\addtolength{\arraycolsep}{-1.0pt}
\end{table}

\section{ Observables Related to the $\Sigma_b \rar \Sigma \ell^+ \ell^-$ Transition}
\subsection{Branching Ratio}
Using the amplitude mentioned above, the differential decay rate
is found as (see also \cite{aliev}) \bea \frac{d\Gamma(z,\hat
s)}{d\hat s~~dz} = \frac{G_F^2\alpha^2_{em} m_{\Sigma_b}}{16384
\pi^5}| V_{tb}V_{ts}^*|^2 v \sqrt{\lambda} \, \Bigg[{\cal
T}_0(\hat s)+{\cal T}_1(\hat s) z +{\cal T}_2(\hat s) z^2\Bigg]~,
\label{ddf} \eea where $\hat s=\frac{q^2}{m_{\Sigma_b}^2}$ and
$z=\cos\theta$ with $\theta$ being the angle between the momenta
of $\Sigma_b$ and $\ell^+$ in the center of mass of leptons,
$\lambda=\lambda(1,r,\hat s)=1+r^2+\hat s^2-2r-2\hat s-2r\hat s$,
$r= m^2_{\Sigma}/m^2_{\Sigma_b}$ and $v=\sqrt{1-\frac{4
m_\ell^2}{q^2}}$.
 The functions, ${\cal T}_0(\hat s)$,  ${\cal T}_1(\hat s)$ and
${\cal T}_2(\hat s)$ are given as
\bea {\cal T}_0(\hat s) \es 32 m_\ell^2
m_{\Sigma_b}^4 \hat s (1+r-\hat s) \ga \vel D_3 \ver^2 +
\vel E_3 \ver^2 \dr \nnb \\
\ar 64 m_\ell^2 m_{\Sigma_b}^3 (1-r-\hat s) \, \mbox{\rm Re} [D_1^\ast
E_3 + D_3
E_1^\ast] \nnb \\
\ar 64 m_{\Sigma_b}^2 \sqrt{r} (6 m_\ell^2 - m_{\Sigma_b}^2 \hat s)
{\rm Re} [D_1^\ast E_1] \nnb \\
\ar 64 m_\ell^2 m_{\Sigma_b}^3 \sqrt{r} \Big( 2 m_{\Sigma_b} \hat s
{\rm Re} [D_3^\ast E_3] + (1 - r + \hat s)
{\rm Re} [D_1^\ast D_3 + E_1^\ast E_3]\Big) \nnb \\
\ar 32 m_{\Sigma_b}^2 (2 m_\ell^2 + m_{\Sigma_b}^2 \hat s) \Big\{ (1
- r + \hat s) m_{\Sigma_b} \sqrt{r} \,
\mbox{\rm Re} [A_1^\ast A_2 + B_1^\ast B_2] \nnb \\
\ek m_{\Sigma_b} (1 - r - \hat s) \, \mbox{\rm Re} [A_1^\ast B_2 +
A_2^\ast B_1] - 2 \sqrt{r} \Big( \mbox{\rm Re} [A_1^\ast B_1] +
m_{\Sigma_b}^2 \hat s \,
\mbox{\rm Re} [A_2^\ast B_2] \Big) \Big\} \nnb \\
\ar 8 m_{\Sigma_b}^2 \Big\{ 4 m_\ell^2 (1 + r - \hat s) +
m_{\Sigma_b}^2 \Big[(1-r)^2 - \hat s^2 \Big]
\Big\} \ga \vel A_1 \ver^2 +  \vel B_1 \ver^2 \dr \nnb \\
\ar 8 m_{\Sigma_b}^4 \Big\{ 4 m_\ell^2 \Big[ \lambda + (1 + r -
\hat s) \hat s \Big] + m_{\Sigma_b}^2 \hat s \Big[(1-r)^2 - \hat s^2 \Big]
\Big\} \ga \vel A_2 \ver^2 +  \vel B_2 \ver^2 \dr \nnb \\
\ek 8 m_{\Sigma_b}^2 \Big\{ 4 m_\ell^2 (1 + r - \hat s) -
m_{\Sigma_b}^2 \Big[(1-r)^2 - \hat s^2 \Big]
\Big\} \ga \vel D_1 \ver^2 +  \vel E_1 \ver^2 \dr \nnb \\
\ar 8 m_{\Sigma_b}^5 \hat s v^2 \Big\{ - 8 m_{\Sigma_b} \hat s \sqrt{r}\,
\mbox{\rm Re} [D_2^\ast E_2] +
4 (1 - r + \hat s) \sqrt{r} \, \mbox{\rm Re}[D_1^\ast D_2+E_1^\ast E_2]\nnb \\
\ek 4 (1 - r - \hat s) \, \mbox{\rm Re}[D_1^\ast E_2+D_2^\ast E_1] +
m_{\Sigma_b} \Big[(1-r)^2 -\hat s^2\Big] \ga \vel D_2 \ver^2 + \vel
E_2 \ver^2\dr \Big\},
\eea
\bea {\cal T}_1(\hat s,1/R) &=& -16
m_{\Sigma_b}^4\s1 v_l \sqrt{\lambda}
\Big\{ 2 Re(A_1^* D_1)-2Re(B_1^* E_1)\nn\\
&+& 2m_{\lb}
Re(B_1^* D_2-B_2^* D_1+A_2^* E_1-A_1^*E_2)\Big\}\nn\\
&+&32 m_{\Sigma_b}^5 \s1~ v_l \sqrt{\lambda} \Big\{
m_{\Sigma_b} (1-r)Re(A_2^* D_2 -B_2^* E_2)\nn\\
&+& \sqrt{r} Re(A_2^* D_1+A_1^* D_2-B_2^*E_1-B_1^* E_2)\Big\},
\eea
 \bea {\cal T}_2(\hat s,1/R)\es - 8 m_{\Sigma_b}^4 v^2 \lambda \ga
\vel A_1 \ver^2 + \vel B_1 \ver^2 + \vel D_1 \ver^2
+ \vel E_1 \ver^2 \dr \nnb \\
\ar 8 m_{\Sigma_b}^6 \hat s v^2 \lambda \Big( \vel A_2 \ver^2 + \vel
B_2 \ver^2 + \vel D_2 \ver^2 + \vel E_2 \ver^2  \Big), \eea
where \bea
\label{nolabel}
A_1 \es - {2 m_b\over q^2} C_7^{eff} \ga f_1^T + g_1^T \dr +
C_9^{eff} \ga f_1-g_1 \dr,\nnb \\
A_2 \es A_1 \ga 1 \rar 2 \dr,\nnb \\
A_3 \es A_1 \ga 1 \rar 3 \dr,\nnb \\
B_i \es A_i \ga g_i \rar - g_i;~g_i^T \rar - g_i^T \dr,\nnb \\
D_1 \es C_{10} \ga f_1-g_1 \dr,\nnb \\
D_2 \es D_1 \ga 1 \rar 2 \dr, \nnb \\
D_3 \es D_1 \ga 1 \rar 3 \dr,\nnb \\
E_i \es D_i \ga g_i \rar - g_i \dr.\eea

To numerically analyze the differential branching ratio with
respect to $q^2$, we perform integral over $z$ in the interval
$z\in[-1,1]$ in Eq. (\ref{ddf}) and take the values of input
parameters as $m_t=(173.5\pm0.6\pm0.8)~GeV$, $m_W=(80.385\pm0.015)~GeV$, $m_Z=(91.1876\pm0.0021)~GeV$, $m_\Sigma = (1192.642\pm0.024)~MeV$,
$m_{\Sigma_{b}} = (5815.5\pm1.8)~ MeV$ and $m_\mu = (105.6583715\pm0.0000035)~MeV$ \cite{char},
$m_b=(4.8\pm0.1)~GeV$,
 $m_c=(1.46\pm0.05)~GeV$ \cite{col},
$| V_{tb}V_{ts}^\ast|=0.041$, $G_F = 1.17 \times 10^{-5}~
GeV^{-2}$ and $\alpha_{em}=\frac{1}{129}$.

The remaining parameters are related to the family non-universal $Z'$
model.
The modifications on the Wilson coefficients in our case are described by the
four parameters, $\vert B^L_{sb}\vert,~\varphi^L_s,
B^L_{\ell\ell}$ and $B^R_{\ell\ell}$. Constraints to $\vert
B^L_{sb}\vert,~\varphi^L_s$ are put fitting the results of the $B^0_s-\bar B^0_s$ mixing observables to recent measurements performed at
Tevatron and LHC \cite{obserbmixing}. These parameters are chosen as $\vert B^L_{sb}\vert
= (0.4 \pm 0.1)\times 10^2$, $\varphi^L_s=(\pm
150)\pm 100$ to maximize the effects of the additional
$Z'$ gauge boson \cite{alievzprime,alievzppol}. Comparing also the theoretical predictions and observational results for the decay channels of $B \rightarrow X_s \mu^+ \mu^-$ \cite{btoxs,btoxssec}, $B \rightarrow K^* \mu^+ \mu^-$ \cite{btokstar,btokstarsec} and $B\rightarrow  \mu^+ \mu^-$ \cite{aajione}, the parameters $B^L_{\ell\ell}=-9.0 \times 10^{-3}$ and $B^R_{\ell\ell}=1.7 \times 10^{-2} $
are obtained \cite{alievzprime,alievzppol}.

We plot the differential branching ratio, forward backward
asymmetry, other asymmetry parameters and polarizations with respect to $q^2$ to
check the sensitivity of these observables to the model parameters mentioned above at muon channel.
We compare predictions of the $Z'$ and SM models when the uncertainties of the form factors are taken into account.
For this aim, first we plot the variations of the
differential branching ratio in terms of $q^2$ in two models in Figure \ref{dbrmuon.eps}. In this figure, the brown-yellow band surrounded by green lines refers to the family non-universal $Z'$ model while the blue band surrounded by the red lines denotes the SM results.
 From
this figure, we obtain the following results:
\begin{figure}[h!]
\centering
\begin{tabular}{cc}
\epsfig{file=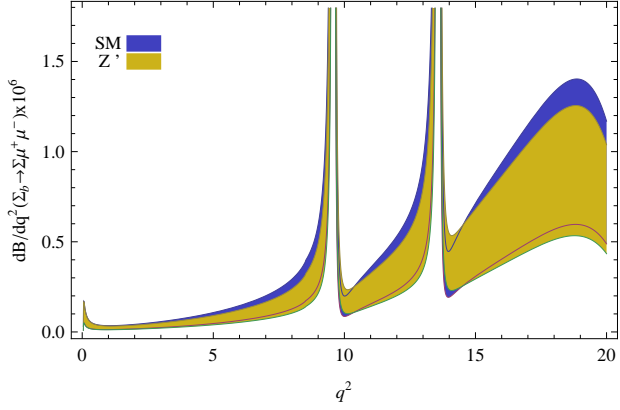,width=0.50\linewidth,clip=}
\end{tabular}
\caption{The  $q^2$ dependence of the differential branching ratio
for the $\Sigma_b \rar \Sigma \mu^+ \mu^-$ decay in family non-universal $Z'$ model as well as in the $SM$.} \label{dbrmuon.eps}
\end{figure}
\begin{itemize}
    \item the SM and $Z'$ bands intersect each other, the discrepancy between the differential branching ratios obtained from two models is small. Hence the differential branching ratio has not essential sensitivity to $Z'$ gauge boson in $\Sigma_b \rar \Sigma \mu^+ \mu^-$ channel. Similar result is obtained in \cite{alievzprime} for  $\Lambda_b \rar \Lambda \mu^+ \mu^-$ decay channel.
     \item The order of differential branching ratio indicates a possibility for this decay mode to be checked at $LHC$ in near future.

\end{itemize}
\subsection{Forward-backward Asymmetry}
\begin{figure}[h!]
\centering
\begin{tabular}{cc}
\epsfig{file=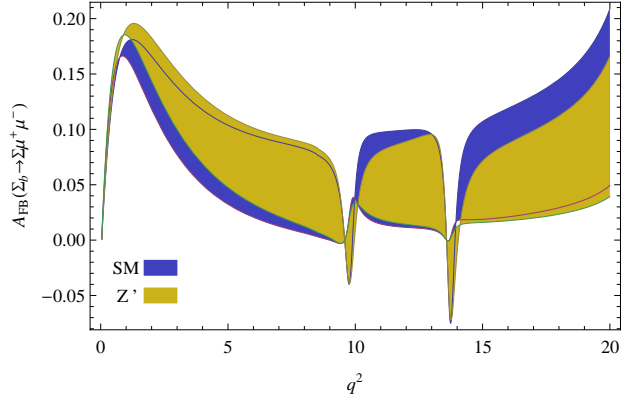,width=0.50\linewidth,clip=}
\end{tabular}
\caption{The  $q^2$ dependence of the forward backward asymmetry
for the $\Sigma_b \rar \Sigma \mu^+ \mu^-$ decay in family non-universal $Z'$ model as well as in the $SM$.} \label{afbmuon.eps}
\end{figure}

One of the most promising tools in detecting the NP effects is the
forward-backward asymmetry of leptons which is defined as
 \bea {\cal A}_{FB} = \frac{N_f-N_b}{N_f+N_b},
 \eea where $N_f$ is the number of events that particle is moving "forward" with respect to
 any chosen direction, and $N_b$ is the number of events that
particle moves to "backward" direction.
 In technique language, the forward-backward asymmetry ${\cal A}_{FB}(\hat s)$ is defined in
terms of the  differential decay rate as \bea {\cal A}_{FB} (\hat
s)= \frac{\ds{\int_0^1\frac{d\Gamma}{d\hat{s}dz}}(z,\hat s)\,dz -
\ds{\int_{-1}^0\frac{d\Gamma}{d\hat{s}dz}}(z,\hat s)\,dz}
{\ds{\int_0^1\frac{d\Gamma}{d\hat{s}dz}}(z,\hat s)\,dz +
\ds{\int_{-1}^0\frac{d\Gamma}{d\hat{s}dz}}(z,\hat s)\,dz}~, \eea
We depict the dependence of ${\cal A}_{FB}$ on $q^2$ in Figure \ref{afbmuon.eps} considering the uncertainties of the form factors.
From this figure we conclude that the effects of $Z'$ model's parameters to the forward backward asymmetry are considerable compared to the differential branching ratio. Although the errors of the form factors partially kill the discrepancies between two models predictions in some regions, they do not sweep the same areas especially at higher values of $q^2$. Similar results have been obtained in $\Lambda_b \rar \Lambda \mu^+ \mu^-$ channel for higher values of $q^2$ \cite{alievzprime}. However small values of $q^2$ the two baryonic channels show different behaviours when we compare the $Z'$ model predictions with those of the SM.
\subsection{Baryon polarization asymmetry parameter - $\alpha_{\Sigma}$}
 Asymmetry parameters characterize the angular dependence of
differential decay width for the cascade decay $\Sigma_b
\rightarrow \Sigma ( \rightarrow a + b) V^* (\rightarrow l^+ l^-
)$ with polarized and unpolarized heavy baryons. In this part,
sensitivity of the baryon polarization asymmetry parameter to new
Wilson coefficients is analyzed. The helicity amplitudes for the
$\Sigma_b \rar\Sigma \ell^+\ell^-$ decay are obtained by analyzing
quasi two body decay $\Sigma_b \rar\Sigma V^\ast$, followed by the
leptonic decay of $V^\ast \rar \ell^+\ell^-$  in \cite{helicity}.
These amplitudes are obtained using helicity amplitude formalism
and polarization density matrix method, demonstrated in
\cite{ampform,polden}. Considering
 \bea \Sigma_b^{1/2 ^+} \rar
\Sigma^{1/2 ^+} \Big( \rar a + b\Big) + V^\ast(\rar \ell^+
\ell^-)~,
 \eea
with $V^\ast$ being off-shell $\gamma$ or $Z$ boson, the
normalized joint angular decay distribution for the two cascade
decay is written as \cite{ampform,2,3,4,5,6} \bea \label{e7121}
\frac{\ds d \Gamma(q^2) }{\ds dq^2 d\!\cos\theta
\,d\!\cos\theta_\Sigma} \es \vel \frac{G_{F} \alpha_{em}}{8
\sqrt{2} \pi} V_{tb} V_{ts}^\ast \ver^2
\frac{\sqrt{\lambda(m_{\Sigma_b}^2,m_\Sigma^2,q^2)}
\sqrt{\lambda(m_\Sigma^2,m_a^2,m_b^2)}} {1024 \pi^3 m_{\Sigma_b}^3
m_\Sigma^2}
 v {\cal B}(\Sigma_b \rar a + b) \vel {\cal M} \ver^2~, \nnb \\
\eea where $\cal M$ is calculated in \cite{helicity}.
 For the leptonic part, in the rest frame of the intermediate
boson, the angle of the anti-lepton with respect to its helicity
axes is shown by $\theta$.
 For the hadronic decay, in the rest frame of the $\Sigma_b$, $\theta_\Sigma$ is the angle of the $a$
momentum with respect to its helicity axes.

\begin{figure}[h!]
\centering
\begin{tabular}{cc}
\epsfig{file=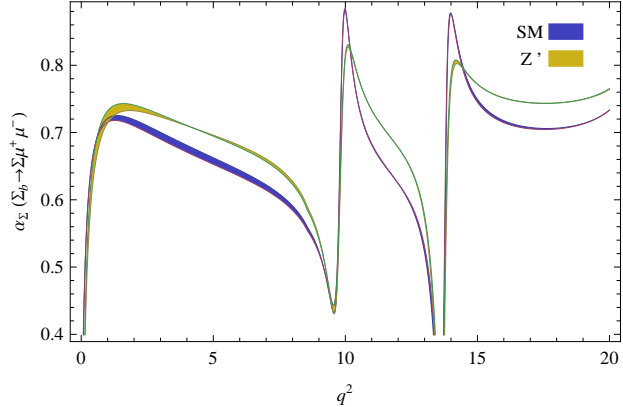,width=0.50\linewidth,clip=}
\end{tabular}
\caption{The $q^2$ dependence of the asymmetry parameter
$\alpha_\Sigma$.} \label{alphalambdamuon.eps}
\end{figure}

The polar angle distribution of $\Sigma \rightarrow a+b$ decay is
obtained by integrating Eq. (\ref{e7121}) with respect to $\theta$
and it gives the differential decay rate in terms of $q^2$ and
$\theta_{\Sigma}$ \cite{helicity} \bea \label{e7126}
\frac{d\Gamma(q^2) }{dq^2\,d\cos\theta_\Sigma} \sim 1 +
\alpha(q^2) \alpha_\Sigma(q^2) \cos\theta_\Sigma (q^2)~, \eea
where the baryon polarization asymmetry parameter
$\alpha_{\Sigma}$ is given as \bea \label{e7134}
\alpha_{\Sigma}(q^2)  \es \frac{8}{3 \Delta(q^2)} \Big\{ 4
m_\ell^2 \vel A_{+1/2,+1} \ver^2 + 2 q^2 \Big( \vel A_{+1/2,+1}
\ver^2 + v^2 \vel B_{+1/2,+1} \ver^2
\Big) \nnb \\
\ek 2 m_\ell^2 \vel A_{+1/2,0} \ver^2 - q^2 \Big( \vel A_{+1/2,0}
\ver^2 + v^2 \vel B_{+1/2,0} \ver^2 \Big)
- 6 m_\ell^2 \vel B_{+1/2,t} \ver^2 \nnb \\
\ek 4 m_\ell^2 \vel A_{-1/2,-1} \ver^2 - 2 q^2 \Big( \vel
A_{-1/2,-1} \ver^2 + v^2 \vel B_{-1/2,-1}
\ver^2\Big) \nnb \\
\ar 2 m_\ell^2 \vel A_{-1/2,0} \ver^2+ q^2 \Big( \vel A_{-1/2,0}
\ver^2 + v^2 \vel B_{-1/2,0} \ver^2 \Big) + 6 m_\ell^2 \vel
B_{-1/2,t} \ver^2\Bigg\}~, \eea and the asymmetry parameter
$\alpha$ is obtained as
 \bea \label{e7127} \alpha(q^2)
\es \frac{8}{3 \Delta(q^2)}\Big\{4 m_\ell^2 \vel A_{+1/2,+1}
\ver^2 + 2 q^2 \Big(\vel A_{+1/2,+1} \ver^2 + v^2 \vel B_{+1/2,+1}
\ver^2
\Big) \nnb \\
\ar 2 m_\ell^2 \vel A_{+1/2,0} \ver^2
+ 6 m_\ell^2 \vel B_{+1/2,t} \ver^2 + q^2 \Big(\vel A_{+1/2,0}
\ver^2 + v^2 \vel B_{+1/2,0} \ver^2 \Big) \nnb \\
\ek 4 m_\ell^2 \vel A_{-1/2,-1} \ver^2 - 2 q^2
\Big(\vel A_{-1/2,-1} \ver^2 + v^2 \vel B_{-1/2,-1} \ver^2 \Big) \nnb \\
\ek q^2 \big(\vel A_{-1/2,0} \ver^2 + v^2 \vel B_{-1/2,0}
\ver^2\Big) - 2 m_\ell^2 \vel A_{-1/2,0} \ver^2 - 6 m_\ell^2 \vel
B_{-1/2,t} \ver^2 \Big\}~, \eea with \bea \label{e7128}
\Delta(q^2)  \es \frac{8}{3}\Big\{4 m_\ell^2 \vel A_{+1/2,+1}
\ver^2 + 2 q^2 \Big(\vel A_{+1/2,+1} \ver^2 + v^2 \vel B_{+1/2,+1}
\ver^2
\Big) \nnb \\
\ar 2 m_\ell^2 \vel A_{+1/2,0} \ver^2
+ 6 m_\ell^2 \vel B_{+1/2,t} \ver^2 + q^2 \Big(\vel A_{+1/2,0}
\ver^2 + v^2 \vel B_{+1/2,0} \ver^2 \Big) \nnb \\
\ar 4 m_\ell^2 \vel
A_{-1/2,-1} \ver^2 + 2 q^2
\Big(\vel A_{-1/2,-1} \ver^2 + v^2 \vel B_{-1/2,-1} \ver^2 \Big) \nnb \\
\ar q^2 \big(\vel A_{-1/2,0} \ver^2 + v^2 \vel B_{-1/2,0}
\ver^2\Big) + 2 m_\ell^2 \vel A_{-1/2,0} \ver^2 + 6 m_\ell^2 \vel
B_{-1/2,t} \ver^2 \Big\}~. \eea The definitions for
$A_{{\lambda_i},\lambda_V}$ and $B_{{\lambda_i},\lambda_V}$ are
given in \cite{helicity} where $\lambda_i$ and $\lambda_V$ are
helicities of lepton pairs and vector boson, respectively. We plot
the dependence of the baryon asymmetry parameter $\alpha_{\Sigma}$
on $q^2$ in Figure \ref{alphalambdamuon.eps}.

From this figure, it is clear that
the family non-universal $Z'$ model's prediction deviates significantly from the SM prediction such that the errors of the form factors can not kill the discrepancies between two model predictions for the baryon asymmetry parameter $\alpha_{\Sigma}$.
\subsection{Polar angle distribution parameters- $\alpha_{\theta}$ and $\beta_{\theta}$}
\begin{figure}[h!] \centering
\begin{tabular}{cc}
\epsfig{file=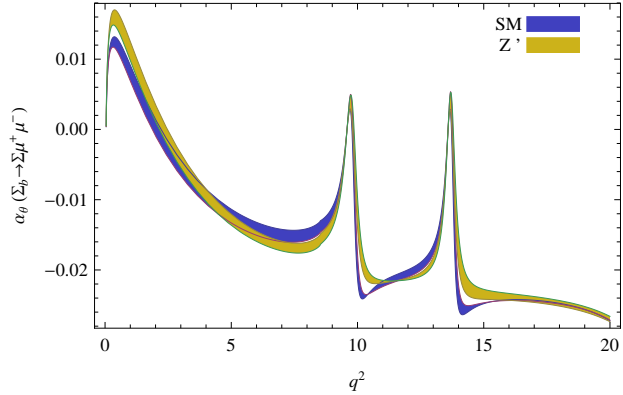,width=0.50\linewidth,clip=}
\end{tabular}
\caption{The $q^2$ dependence of the asymmetry parameter
$\alpha_\theta$.} \label{alphathetam.eps}
\end{figure}

\begin{figure}[h!]
\centering
\begin{tabular}{cc}
\epsfig{file=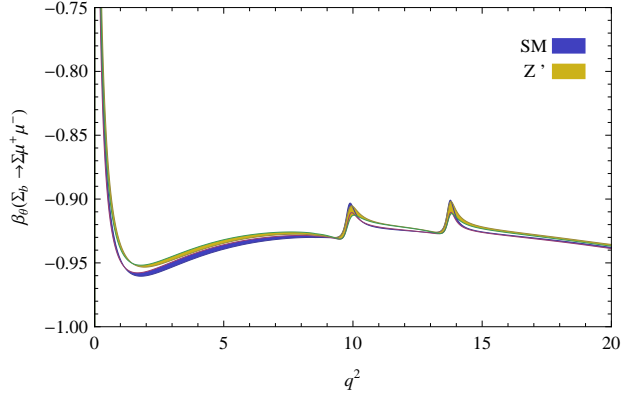,width=0.50\linewidth,clip=}
\end{tabular}
\caption{The $q^2$ dependence of the asymmetry parameter
$\beta_\theta$.} \label{alphabetamuon.eps}
\end{figure}

The polar angle distribution of $V^* \rightarrow l^+l^-$ decay is
obtained by integrating Eq. (\ref{e7121}) with respect to
$\theta_{\Sigma}$ which gives the differential decay rate in terms
of $q^2$ and $\theta$
 \bea \label{e7129} \frac{d\Gamma(q^2)
}{dq^2\,d\cos\theta} \sim 1 + 2 \alpha_\theta \cos\theta +
\beta_\theta \cos^2\theta~, \eea where \bea \label{e7130}
\alpha_\theta(q^2)  \es \frac{1}{\Delta_1(q^2)} 2 v q^2 \mbox{\rm
Re} \Big[ A_{+1/2,+1} B_{+1/2,+1}^\ast - A_{-1/2,-1}
B_{-1/2,-1}^\ast \Big]~, \eea and \bea \label{e7131}
\beta_\theta(q^2)  \es \frac{1}{\Delta_1(q^2)} \Big\{ - 4 m_\ell^2
\vel A_{+1/2,+1} \ver^2 +
q^2 \Big( \vel A_{+1/2,+1} \ver^2 + v^2 \vel B_{+1/2,+1} \ver^2 \Big) \nnb \\
\ar 4 m_\ell^2 \vel A_{+1/2,0} \ver^2 -
q^2 \Big( \vel A_{+1/2,0} \ver^2 + v^2 \vel B_{+1/2,0} \ver^2 \Big) \nnb \\
\ek 4 m_\ell^2 \vel A_{-1/2,-1} \ver^2 + q^2
\Big( \vel A_{-1/2,-1} \ver^2 + v^2 \vel B_{-1/2,-1} \ver^2 \Big) \nnb \\
\ar 4 m_\ell^2 \vel A_{-1/2,0} \ver^2 -
q^2 \Big( \vel A_{-1/2,0} \ver^2 + v^2 \vel B_{-1/2,0} \ver^2 \Big)
\Big\}~,
\eea
with
\bea
\label{e7132}
\Delta_1(q^2)  \es 4 m_\ell^2  \vel A_{+1/2,+1} \ver^2 +
q^2 \Big( \vel A_{+1/2,+1} \ver^2 + v^2 \vel B_{+1/2,+1} \ver^2 \Big) \nnb \\
\ar 4 m_\ell^2 \vel B_{+1/2,t} \ver^2
+ q^2 \Big( \vel A_{+1/2,0} \ver^2 + v^2 \vel B_{+1/2,0} \ver^2 \Big) \nnb \\
\ar 4 m_\ell^2 \vel A_{-1/2,-1} \ver^2 + q^2
\Big( \vel A_{-1/2,-1} \ver^2 + v^2 \vel B_{-1/2,-1} \ver^2 \Big) \nnb \\
\ar  4 m_\ell^2 \vel B_{-1/2,t} \ver^2 + q^2 \Big( \vel A_{-1/2,0}
\ver^2 + v^2 \vel B_{-1/2,0} \ver^2 \Big)~. \eea We plot the
dependence of the polar angle distribution parameters
$\alpha_{\theta}$ and  $\beta_{\theta}$ on $q^2$ in Figures \ref{alphathetam.eps}
and \ref{alphabetamuon.eps}, respectively. From these figures, we read that

\begin{itemize}
\item as far as the $\alpha_{\theta}$ is considered we see a considerable discrepancy between the two model's predictions. In the case of  $\Lambda_b \rar \Lambda \mu^+ \mu^-$ this parameter also shows to be very sensitive to $Z'$ gauge boson. However in our case this sensitivity is relatively small.
\item Similar to the $\Lambda_b \rar \Lambda \mu^+ \mu^-$ channel
considered in \cite{alievzprime}, the asymmetry parameter
$\beta_{\theta}$ in $Z^{\prime}$ model gives roughly the same
result as the SM.
\end{itemize}
\newpage
\subsection{Double Lepton Polarizations}
\begin{figure}[h!]
\centering
\begin{tabular}{cc}
\epsfig{file=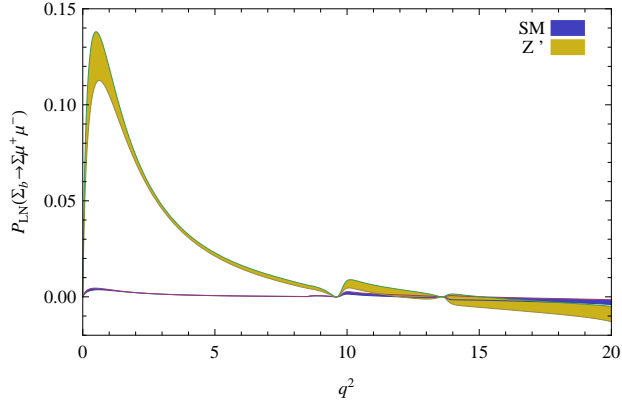,width=0.5\linewidth,clip=}
\end{tabular}
\caption{The $q^2$ dependence of Longitudinal-normal (LN)
polarization of leptons} \label{PLNmuon.eps}
\end{figure}

\begin{figure}[h!]
\centering
\begin{tabular}{cc}
\epsfig{file=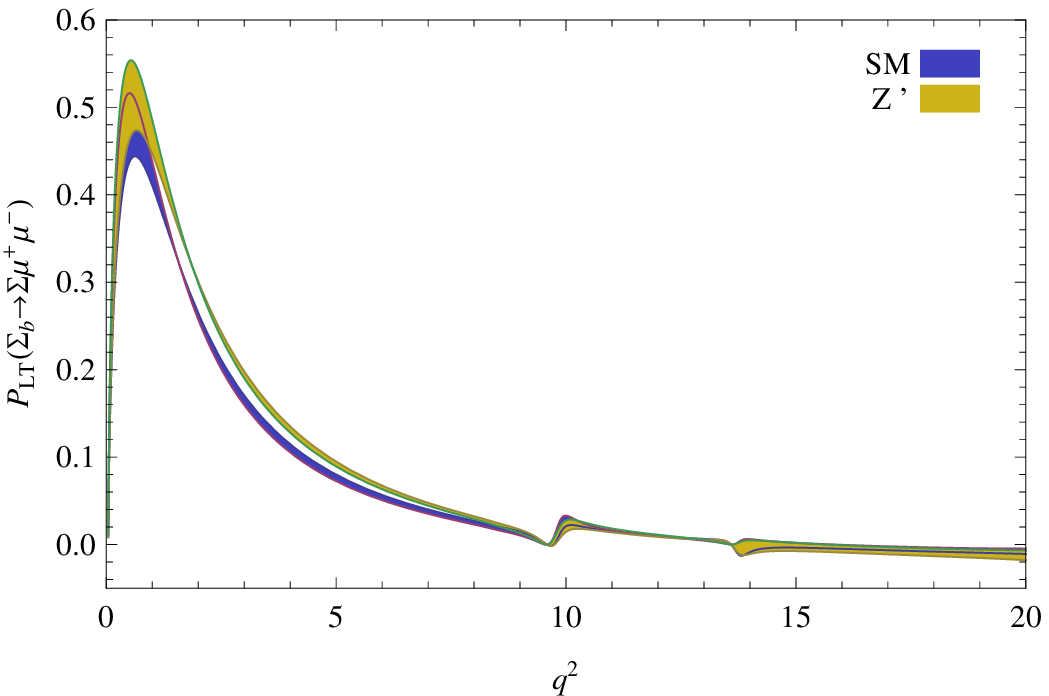,width=0.5\linewidth,clip=}
\end{tabular}
\caption{The same as figure 6, but for LT polarization}
\label{PLTmuon.eps}
\end{figure}

\begin{figure}[h!]
\centering
\begin{tabular}{cc}
\epsfig{file=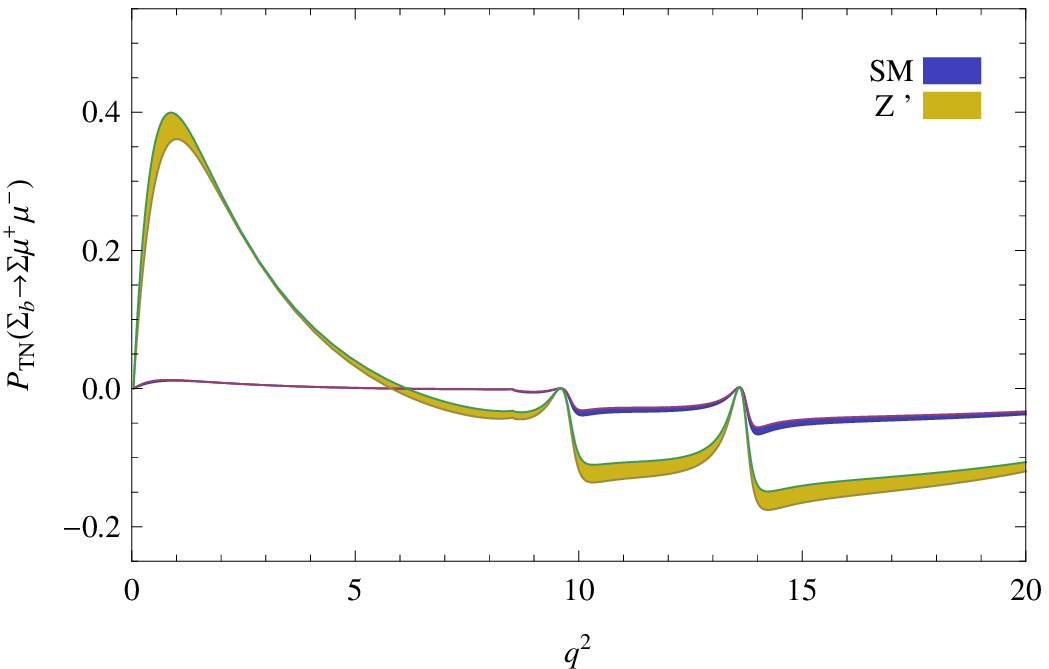,width=0.5\linewidth,clip=}
\end{tabular}
\caption{The same as figure 6, but for TN polarization}
\label{PTNmuon.eps}
\end{figure}

\begin{figure}[h!]
\centering
\begin{tabular}{cc}
\epsfig{file=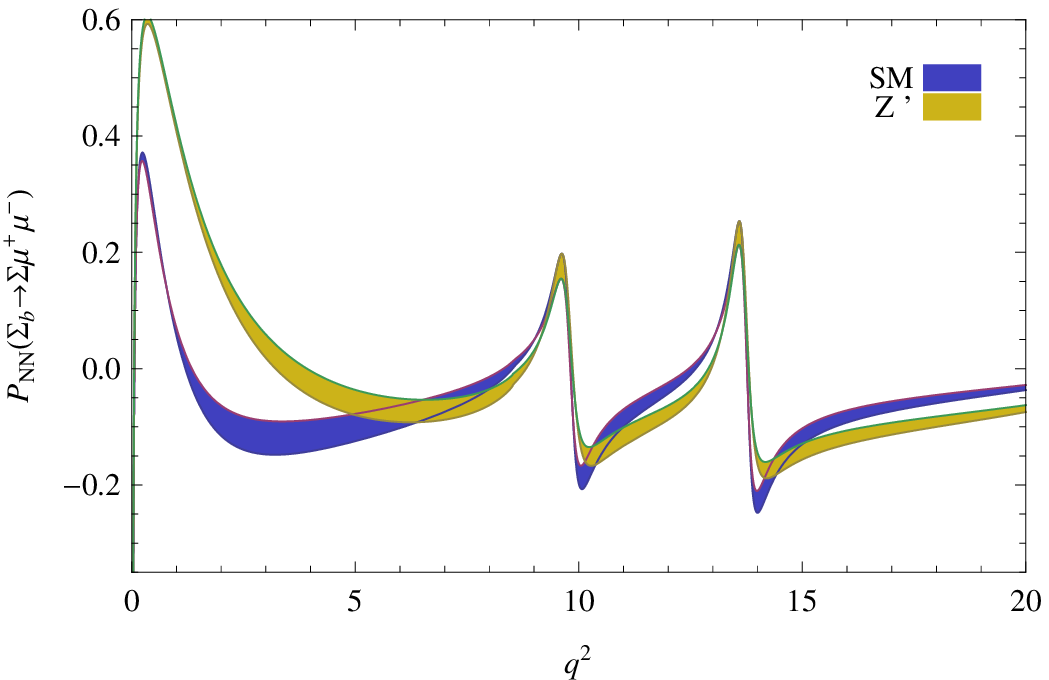,width=0.5\linewidth,clip=}
\end{tabular}
\caption{The same as figure 6, but for NN polarization}
\label{PNNmuon.eps}
\end{figure}

\begin{figure}[h!]
\centering
\begin{tabular}{cc}
\epsfig{file=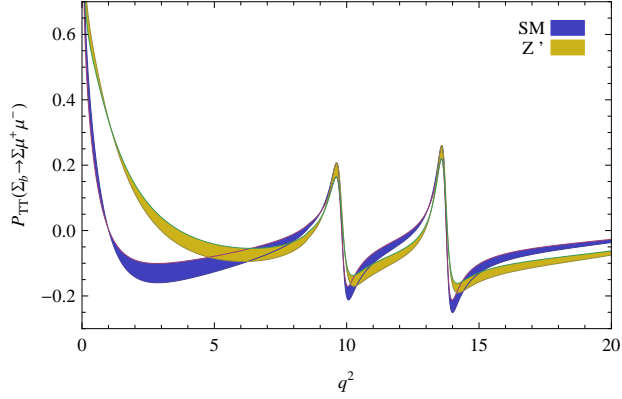,width=0.5\linewidth,clip=}
\end{tabular}
\caption{The same as figure 6, but for TT polarization}
\label{PTTm.eps}
\end{figure}

To analyze the double lepton polarizations, we need to define the
following orthogonal unit vectors $s_i^{\pm\mu}$ with $i=L,T$ or
$N$ in the rest frame of double leptons (see for instance
\cite{AlievSirvanli,alievbashiry,Bensalem} for detailed
information)
 \bea \label{e6616} s^{-\mu}_L \es \ga
0,\vec{e}_L^{\,-}\dr =
\ga 0,\frac{\vec{p}_-}{\vel\vec{p}_- \ver}\dr~, \nnb \\
s^{-\mu}_N \es \ga 0,\vec{e}_N^{\,-}\dr = \ga
0,\frac{\vec{p}_\Sigma\times
\vec{p}_-}{\vel \vec{p}_\Sigma\times \vec{p}_- \ver}\dr~, \nnb \\
s^{-\mu}_T \es \ga 0,\vec{e}_T^{\,-}\dr = \ga 0,\vec{e}_N^{\,-}
\times \vec{e}_L^{\,-} \dr~, \nnb \\
s^{+\mu}_L \es \ga 0,\vec{e}_L^{\,+}\dr =
\ga 0,\frac{\vec{p}_+}{\vel\vec{p}_+ \ver}\dr~, \nnb \\
s^{+\mu}_N \es \ga 0,\vec{e}_N^{\,+}\dr = \ga
0,\frac{\vec{p}_\Sigma\times
\vec{p}_+}{\vel \vec{p}_\Sigma\times \vec{p}_+ \ver}\dr~, \nnb \\
s^{+\mu}_T \es \ga 0,\vec{e}_T^{\,+}\dr = \ga 0,\vec{e}_N^{\,+}
\times \vec{e}_L^{\,+}\dr~, \eea where $\vec{p}_\pm$ and
$\vec{p}_\Sigma$ are the three-momenta of the leptons $\ell^\pm$
and $\Sigma$ baryon. The symbols L, T and N denote longitudinal,
transverse, and normal polarizations, respectively. Using the
Lorentz boost, these unit vectors are transformed from the rest
frame of the leptons into the center of mass (CM) frame of them
along the longitudinal direction. As a result, we get
 \bea \label{e6617} \ga s^{\mp\mu}_L \dr_{CM} \es \ga
\frac{\vel\vec{p}_\mp \ver}{m_\ell}~, \frac{E_\ell~
\vec{p}_\mp}{m_\ell \vel\vec{p}_\mp \ver}\dr~, \eea where
$\vec{p}_+ = - \vec{p}_-$ and  $E_\ell$ and $m_\ell$ are the
energy and mass of leptons in the CM frame, respectively. Under
the above transformation, the other two unit vectors,
$s_N^{\pm\mu}$, $s_T^{\pm\mu}$ remain unchanged. The double lepton
polarizations are given as \cite{AlievSirvanli,alievbashiry,Bensalem} \bea
\label{e111213} P_{ij}(\hat{s}) \es \frac{ \Big( \ds
\frac{d\Gamma(\vec{s}^{\,\,-}_i,\vec{s}^{\,\,+}_j)}{d \hat{s}} -
      \ds \frac{d\Gamma(-\vec{s}^{\,\,-}_i,\vec{s}^{\,\,+}_j)}{d \hat{s}} \Big) -
\Big( \ds \frac{d\Gamma(\vec{s}^{\,\,-}_i,-\vec{s}^{\,\,+}_j)}{d \hat{s}} -
      \ds \frac{d\Gamma(-\vec{s}^{\,\,-}_i,-\vec{s}^{\,\,+}_j)}{d \hat{s}} \Big)
     }
     {
\Big( \ds \frac{d\Gamma(\vec{s}^{\,\,-}_i,\vec{s}^{\,\,+}_j)}{d \hat{s}} +
      \ds \frac{d\Gamma(-\vec{s}^{\,\,-}_i,\vec{s}^{\,\,+}_j)}{d \hat{s}} \Big) +
\Big( \ds \frac{d\Gamma(\vec{s}^{\,\,-}_i,-\vec{s}^{\,\,+}_j)}{d \hat{s}} +
      \ds \frac{d\Gamma(-\vec{s}^{\,\,-}_i,-\vec{s}^{\,\,+}_j)}{d \hat{s}} \Big)
     }~. \nnb \\ \nnb \\
     \eea
The first (second) subindex of $P_{ij}$ represents polarization of
lepton (anti-lepton). Using the above definitions, some double
lepton polarizations are obtained as \bea \label{e111215}
P_{LN}(\hat{s}) \es - P_{NL}(\hat{s}) = \frac{16 \pi
m_{\Sigma_b}^4 \hat{m}_\ell
\sqrt{\lambda}}{\Delta^{\prime}(\hat{s}) \sqrt{\hat{s}}} \mbox{\rm
Im} \Bigg\{
(1-\hat{r})
(A_1^\ast D_1 + B_1^\ast E_1) \nnb \\
\ar m_{\Sigma_b}
 \hat{s} (A_1^\ast E_3 - A_2^\ast E_1 + B_1^\ast D_3
-B_2^\ast D_1) \nnb \\
\ar m_{\Sigma_b}
 \sqrt{\hat{r}} \hat{s}
(A_1^\ast D_3 + A_2^\ast D_1 +B_1^\ast E_3 + B_2^\ast E_1)
- m_{\Sigma_b}^2 \hat{s}^2 (B_2^\ast E_3 + A_2^\ast D_3)
\Bigg\}~,
\eea
 \bea \label{e111217} P_{LT}(\hat{s}) \es \frac{16 \pi
m_{\Sigma_b}^4 \hat{m}_\ell \sqrt{\lambda}
v}{\Delta^{\prime}(\hat{s}) \sqrt{\hat{s}}} \mbox{\rm Re} \Bigg\{
(1-\hat{r}) \Big( \vel D_1 \ver^2 + \vel E_1 \ver^2 \Big)
- \hat{s} \Big(A_1 D_1^\ast - B_1 E_1^\ast
\Big) \nnb \\
\ek m_{\Sigma_b} \hat{s} \Big[ B_1 D_2^\ast + (A_2 + D_2 -D_3)
E_1^\ast -  A_1 E_2^\ast -(B_2-E_2+E_3) D_1^\ast \Big]
\nnb \\
\ar m_{\Sigma_b}^2 \hat{s} (1-\hat{r})
(A_2 D_2^\ast - B_2 E_2^\ast) \nnb \\
\ar m_{\Sigma_b}
 \sqrt{\hat{r}} \hat{s}
\Big[ A_1 D_2^\ast + (A_2 + D_2 +D_3) D_1^\ast - B_1 E_2^\ast -
(B_2 - E_2 - E_3) E_1^\ast \Big] \nnb \\
\ek m_{\Sigma_b}^2 \hat{s}^2 (D_2 D_3^\ast + E_2 E_3^\ast )
\Bigg\}~,
\eea
\bea \label{e111219} P_{NT}(\hat{s}) \es - P_{TN} = \frac{64
m_{\Sigma_b}^4 \lambda v}{3 \Delta^{\prime}(\hat{s})} \mbox{\rm
Im} \Bigg\{
(A_1 D_1^\ast +B_1 E_1^\ast)
+ m_{\Sigma_b}^2\hat{s} (A_2^\ast D_2 + B_2^\ast E_2)
\Bigg\}~,
\eea
\bea \label{e111221} P_{NN}(\hat{s}) \es \frac{32
m_{\Sigma_b}^4}{3 \hat{s} \Delta^{\prime}(\hat{s})} \mbox{\rm Re}
\Bigg\{
24 \hat{m}_\ell^2 \sqrt{\hat{r}} \hat{s} ( A_1 B_1^\ast +
D_1 E_1^\ast ) - 12 m_{\Sigma_b} \hat{m}_\ell^2
\sqrt{\hat{r}} \hat{s}
(1-\hat{r} +\hat{s}) (A_1 A_2^\ast + B_1 B_2^\ast) \nnb \\
\ar 6 m_{\Sigma_b} \hat{m}_\ell^2 \hat{s} \Big[ m_{\Sigma_b}
\hat{s} (1+\hat{r}-\hat{s}) \Big(\vel D_3 \ver^2 + \vel E_3
\ver^2 \Big) + 2 \sqrt{\hat{r}} (1-\hat{r}+\hat{s})
(D_1 D_3^\ast + E_1 E_3^\ast)\Big] \nnb \\
\ar 12 m_{\Sigma_b} \hat{m}_\ell^2 \hat{s}
(1-\hat{r}-\hat{s})
(A_1 B_2^\ast + A_2 B_1 ^\ast + D_1 E_3^\ast + D_3 E_1^\ast) \nnb \\
\ek [ \lambda \hat{s} + 2 \hat{m}_\ell^2 (1 + \hat{r}^2 - 2
\hat{r} + \hat{r} \hat{s} + \hat{s} - 2 \hat{s}^2) ]
\Big( \vel A_1 \ver^2 + \vel B_1 \ver^2 - \vel D_1 \ver^2 -
\vel E_1 \ver^2 \Big) \nnb \\
\ar 24 m_{\Sigma_b}^2 \hat{m}_\ell^2 \sqrt{\hat{r}}
\hat{s}^2 (A_2 B_2^\ast + D_3 E_3^\ast)
- m_{\Sigma_b}^2 \lambda \hat{s}^2 v^2
\Big( \vel D_2 \ver^2 + \vel E_2 \ver^2 \Big) \nnb \\
\ar m_{\Sigma_b}^2 \hat{s} \{ \lambda \hat{s} - 2 \hat{m}_\ell^2
[2 (1+ \hat{r}^2) - \hat{s} (1+\hat{s}) - \hat{r}
(4+\hat{s})]\} \Big( \vel A_2 \ver^2 + \vel B_2 \ver^2 \Big)
\Bigg\}~,
 \eea
 \bea \label{e111222} P_{TT}(\hat{s}) \es \frac{32
m_{\Sigma_b}^4}{3 \hat{s} \Delta^{\prime}(\hat{s})} \mbox{\rm Re}
\Bigg\{
- 24 \hat{m}_\ell^2 \sqrt{\hat{r}} \hat{s} ( A_1 B_1^\ast +
D_1 E_1^\ast )
- 12 m_{\Sigma_b} \hat{m}_\ell^2 \sqrt{\hat{r}} \hat{s}
(1-\hat{r} +\hat{s}) (D_1 D_3^\ast + E_1 E_3^\ast) \nnb \\
\ek 24 m_{\Sigma_b}^2 \hat{m}_\ell^2 \sqrt{\hat{r}}
\hat{s}^2
( A_2 B_2^\ast + D_3 E_3^\ast ) \nnb \\
\ek 6 m_{\Sigma_b} \hat{m}_\ell^2 \hat{s} \Big[ m_{\Sigma_b}
\hat{s} (1+\hat{r}-\hat{s}) \Big(\vel D_3 \ver^2 + \vel E_3
\ver^2 \Big) - 2 \sqrt{\hat{r}} (1-\hat{r}+\hat{s})
(A_1 A_2^\ast + B_1 B_2^\ast)\Big] \nnb \\
\ek 12 m_{\Sigma_b} \hat{m}_\ell^2 \hat{s}
(1-\hat{r}-\hat{s})
(A_1 B_2^\ast + A_2 B_1 ^\ast + D_1 E_3^\ast + D_3 E_1^\ast) \nnb \\
\ek [ \lambda \hat{s} - 2 \hat{m}_\ell^2 (1 + \hat{r}^2 - 2
\hat{r} + \hat{r} \hat{s} + \hat{s} - 2 \hat{s}^2) ]
\Big( \vel A_1 \ver^2 + \vel B_1 \ver^2 \Big) \nnb \\
\ar m_{\Sigma_b}^2 \hat{s} \{ \lambda \hat{s} + \hat{m}_\ell^2 [4
(1- \hat{r})^2 - 2 \hat{s} (1+\hat{r}) - 2
\hat{s}^2 ]\}
\Big( \vel A_2 \ver^2 + \vel B_2 \ver^2 \Big) \nnb \\
\ar \{ \lambda \hat{s} - 2 \hat{m}_\ell^2 [5 (1- \hat{r})^2
- 7 \hat{s} (1+\hat{r}) + 2 \hat{s}^2 ]\}
\Big( \vel D_1 \ver^2 + \vel E_1 \ver^2 \Big) \nnb \\
\ek m_{\Sigma_b}^2 \lambda \hat{s}^2 v^2 \Big( \vel D_2 \ver^2 +
\vel E_2 \ver^2 \Big)
\Bigg\}~, \eea
 where \bea \Delta^{\prime}(\hat s)\es{{\cal T}_0(\hat s)
+\frac{1}{3} {\cal T}_2(\hat s)}.
 \eea
  We plot the dependence of the double lepton
polarizations on $q^2$ in Figures \ref{PLNmuon.eps}-\ref{PTTm.eps}.

From these figures, we observe that
\begin{itemize}
\item in general, there are considerable discrepancies between the
$Z^\prime$ model and the SM predictions on the double lepton
polarization asymmetries under consideration except for the $P_{LT}$.
\item The big deviations of
the $Z^{\prime}$ model's results from the SM predictions appear at
lower values of $q^2$. For the $P_{NN}$ and $P_{TT}$, the NP parameters are more effective at lower values of
$q^2$.
\item The maximum discrepancy between two model's predictions belongs to the $P_{LN}$
and $P_{TN}$.

\item Compared to the $\Lambda_b \rar \Lambda \mu^+ \mu^-$ decay channel \cite{alievzppol}, the double lepton polarization asymmetries are more sensitive to the $Z'$ gauge boson in  $\Sigma_b \rar \Sigma \mu^+ \mu^-$ decay channel.
\item Determination of zero points of the double lepton
polarization asymmetries in the experiment and comparison of the obtained results
with the theoretical predictions can give valuable information
about the existence of the $Z^{\prime}$ boson.
\end{itemize}

\section{Conclusion}
In the present study, we analyzed the $\Sigma_b \rar \Sigma \mu^+
\mu^-$ decay channel in both the SM and $Z^\prime$ model considering the errors of form factors. We discussed the sensitivity of the differential branching ratio, forward-backward asymmetry as well as some asymmetry parameters and polarization asymmetries defining the considered decay channel to $Z'$ gauge boson. Our results on the considered physical quantities overall depict considerable discrepancies between the
$Z^{\prime}$ and SM model's predictions. In the case of the differential branching ratio the discrepancy between two model's predictions is small such that the uncertainties of the form factors roughly kill the difference. The maximum discrepancies belong to the double lepton polarization asymmetries and asymmetry parameter $\alpha_{\Sigma}$ which are quite sensitive to $Z'$ gauge boson. The discrepancies between two model's predictions on some physical observables can be considered as signals for existing the extra $Z^{\prime}$ gauge
boson. The order of the differential branching ratio indicates
that this channel can be studied at LHC in near future. Any
measurement on the physical quantities considered in the present
work and comparison of the obtained data with the theoretical
predictions can give useful information not only about the
existence of the $Z^{\prime}$ gauge boson but also about the
nature of participating baryons, $\Sigma_b$ and $\Sigma$. We also compared our results with the results of the considered observables for $\Lambda_b \rar \Lambda \mu^+ \mu^-$ decay channel \cite{alievzprime,alievzppol} which is in agenda of different experiments
nowadays. The study of different FCNC channels can provide us with
more data which may help us in searching for $Z^{\prime}$ gauge
boson as new physics effect.
\section{Acknowledgement}
N. Kat{\i}rc{\i} thanks Bo\u gazi\c ci university for the financial support provided by the scientific research fund   with  project no: 7128.


\begin{thebibliography}{99}
\bibitem{cdfobs} T. Aaltonen et al. [CDF Collaboration], Phys. Rev. Lett. 107, 201802 (2011); arXiv: 1107.3753[hep-ex].
\bibitem{lhcb} Our personal communications with Yasmine Sara Amhis from LHCb Collaboration.
\bibitem{formfactor} K. Azizi, M. Bayar, A. Ozpineci, Y. Sarac and H. Sundu, Phys. Rev. D 85, 016002 (2012); arXiv: 1112.5147[hep-ph].
\bibitem{ealti} F. G\"{u}rsey, M. Serdaro\u glu, Lett. Nuo. Cim. 21, (1978).
\bibitem{buchalla} G. Buchalla, G. Burdman, C. T. Hill and D. Kominis, Phys. Rev. D 53, 5185 (1996); arXiv: hep-ph/9510376.
\bibitem{nardi} E. Nardi, Phys. Rev. D 48, 1240 (1993); arXiv: hep-ph/9209223.
\bibitem{bernabeu} J. Bernabeu, E. Nardi and D. Tommasini, Nucl. Phys. B 409, 69 (1993); arXiv: hep-ph/9306251.
\bibitem{bargerv} V. Barger, M. Berger and R. J. Phillips, Phys. Rev. D 52, 1663 (1995); arXiv: hep-ph/9503204.
\bibitem{theoz} E. Eichten, I. Hinchliffe, K. D. Lane and C. Quigg,  Rev. Mod. Phys. 56, 579 (1984).
\bibitem{zp} P. Langacker and M. Plumacher, Phys. Rev. D 62, 013006 (2000); arXiv: hep-ph/0001204.
\bibitem{lepto} J. L. Lopez and D. V. Nanopoulos, Phys. Rev. D 55, 397 (1997); arXiv: hep-ph/9605359.
\bibitem{sirvanli} B. B. Sirvanli, Mod. Phys. Lett. A 23, 347 (2008); arXiv: hep-ph/0701173.
\bibitem{leike} A. Leike, Phys. Rept. 317, 143 (1999); arXiv: hep-ph/9805494.
\bibitem{nonunibir} T. K. Kuo and N. Nakagawa, Phys. Rev. D 31, 1161 (1985); Phys. Rev. D 32, (1985) 306.
\bibitem{nonuniiki} K. K. Gan, Phys. Lett. B 209, 95 (1988).
\bibitem{nonuniuc} E. Nardi, Phys. Rev. D 48, 1240 (1993); arXiv: hep-ph/9211246.
\bibitem{nonunid} B. Holdom, Phys. Lett. B 339, 114 (1994); arXiv: hep-ph/9407311.
\bibitem{nonunib} X. Zhang and B. L. Young, Phys. Rev. D 51, 6584 (1995).
\bibitem{nonunia} B. Holdom and M. V. Ramana, Phys. Lett. B 365, 309 (1996); arXiv: hep-ph/9509272.
\bibitem{nonu} S. Chaudhuri, S.W. Chung, G. Hockney and J. Lykken, Nucl. Phys. B 456, 89 (1995).
\bibitem{nonuiki} G. Cleaver, M. Cvetic, J. R. Espinosa, L. Everett and P. Langacker, Nucl. Phys. B 525, 3 (1998); arXiv: hep-th/9711178.
\bibitem{stu} Y. Zhang, Z. Cai; arXiv: 1106.0163[hep-ph].
\bibitem{ew} A. Kundu, Phys. Lett. B 370, 135 (1996); arXiv: hep-ph/9504417.
\bibitem{erler} J. Erler, P. Langacker, Phys. Rev. Lett. 84, 212 (2000); arXiv: hep-ph/9910315.
\bibitem{fitb} C. Caso et al., Eur. Phys. J. C3, 1 (1998).
\bibitem{paul} P. Langacker, Rev. Mod. Phys. 81, 1199 (2008); arXiv: 0801.1345[hep-ph].
\bibitem{const1} K. Cheung, C. W. Chiang, N. G. Deshpande and J. Jiang, Phys. Lett. B 652, 285 (2007); arXiv: hep-ph/0604223.
\bibitem{constiki} X. G. He and G. Valencia, Phys. Rev. D 74, 013011 (2006); arXiv: hep-ph/0605202.
\bibitem{cdf} F. Abe et al. [CDF collaboration], Phys. Rev. Lett. 79, 2192 (1997).
\bibitem{directsearch} A. Abulencia et al. [CDF collaboration], Phys. Rev. Lett. 96, 211801 (2006); arXiv: hep-ex/0602045.
\bibitem{tevatron} M. Carena, A. Daleo, B. A. Dobrescu, T. M. P. Tait, Phys. Rev. D 70, 093009 (2004); arXiv: hep-ph/0408098.
\bibitem{giri} A. K. Giri, R. Mohanta, Eur. Phys. J. C 45, 151 (2006); arXiv: hep-ph/0510171.
\bibitem{alievzprime} T. M. Aliev, M. Savci, Nuc. Phys. B, 863, 398, (2012), arXiv:1202.0398[hep-ph].
\bibitem{alievzppol} T. M. Aliev, M. Savci, Phys. Lett. B, 718, 566, (2012), arXiv:1202.5444[hep-ph].
\bibitem{zpB2} C. H. Chen and H. Hatanaka, Phys. Rev. D 73, 075003 (2006); arXiv: hep-ph/0602140.
\bibitem{barger} V. Barger, C. W. Chiang, P. Langacker, H. S. Lee, Phys. Lett. B 580, 186 (2004); arXiv: hep-ph/0310073 ; ibid B 598, 218 (2004); arXiv: hep-ph/0406126; V. Barger, L. Everett, J. Jiang, P. Langacker, T. Liu, C. E. M. Wagner, Phys. Rev. D 80, 055008 (2009); arXiv: hep-ph/0902.4507.
\bibitem{zpB3} C. W. Chiang, N. G. Deshpande, J. Jiang, JHEP 0608, 075 (2006); arXiv: hep-ph/0606122.
\bibitem{che} K. Cheung, C. W. Chiang, N. G. Deshpande, J. Jiang, Phys. Lett. B 652, 285 (2007); arXiv: hep-ph/0604223.
\bibitem{mohanta} R. Mohanta and A. K. Giri, Phys. Rev. D 79, 057902 (2009); arXiv: 0812.1842[hep-ph].
\bibitem{hua} J. Hua, C. S. Kim and Y. Li, Phys. Lett. B 690, 508 (2010); arXiv: 1002.2532[hep-ph]; Q. Chang, Nucl. Phys. B 845, 179 (2011); arXiv: 1101.1272[hep-ph].
\bibitem{zpB4} Q. Chang, X. Q. Li and Y. D. Yang, JHEP 0905, 056 (2009); arXiv: 0903.0275[hep-ph].
\bibitem{li} Y. Li, J. Hua and K. C. Yang, JHEP 1002, 082 (2010); arXiv: 0907.4408[hep-ph]; Q. Chang, X. Q. Li and Y. D. Yang, JHEP 1004, 052 (2010); arXiv: 1002.2758[hep-ph].
\bibitem{wang} S. W. Wang, G. L. Sun, X. Q. Yang and J. S. Huang, Eur. Phys. J. C 72, 1852 (2012).
\bibitem{bk2} R. H. Li, C. D. La, W. Wang, Phys. Rev. D 83, 034034 (2011).
\bibitem{bk1} Y. Li, J. Hua, K. C. Yang, Eur. Phys. J. C 71, 1775 (2011); arXiv: 1107.0630[hep-ph].
\bibitem{bk0} Y. Li, X. J. Fan, J. Hua, E. L. Wang, Phys. Rev. D 85, 074010 (2012); arXiv: 1111.7153[hep-ph].
\bibitem{btopi} Q. Chang, X. Q. Li, Y. D. Yang, Int. J. Mod. Phys. A 26, 1273 (2011); arXiv: 1003.6051[hep-ph].
 \bibitem{obserbmixing} X. Qiang Li, Y. Min Li, G. R. Liu, F. Su, (2012); arXiv:1204.5250 [hep-ph].
\bibitem{abazov} V. Abazov, et al., D0 Collaboration, D0 Conference note, 6098-CONF.
\bibitem{aaltonen} T. Aaltonen, et al., CDF Collaboration, Phys. Rev. D 85 072002, (2012).
\bibitem{aajione} R. Aaij, et al., LHCb Collaboration, Phys. Rev. Lett. 108 (2012) 101803; arXiv:1203.4493 [hep-ph].
\bibitem{aajisec} R. Aaij, et al., LHCb Collaboration, Phys. Lett. B 707 497, (2012).
\bibitem{breitwigner} G. Buchalla, A. J. Buras, M. E. Lautenbacher, Rev. Mod. Phys. 68, 1125 (1996); arXiv: hep-ph/9512380.
\bibitem{bobeth} G. Bobeth, A. J. Buras, F. Kr\"{u}ger, J. Urban, Nucl. Phys. B 630, 87 (2002); arXiv: hep-ph/0112305.
\bibitem{altmann} W. Altmannshofer, P. Ball, A. Bharucha, A. Buras, D. M. Straub, M. Wick, JHEP 0901, 019 (2009); arXiv: 0811.1214[hep-ph].
\bibitem{ghin} A. Ghinculov, T. Hurth, G. Isidori, Y. P. Yao, Nucl. Phys. B 685, 351 (2004); arXiv: hep-ph/0312128.
\bibitem{misiak} K.G. Chetyrkin, M. Misiak, M. Munz, Phys. Lett. B 400, 206 (1997).
\bibitem{cyedi} A. J. Buras, M. Misiak, M. Muenz and S. Pokorski, Nucl. Phys. B 424, 374 (1994); arXiv: hep-ph/9311345.
\bibitem{cdokuz} M. Misiak, Nucl. Phys. B  393, 23 (1993); Erratum ibid  B  439, 161 (1995).
\bibitem{cdok} A.J. Buras, M. Muenz, Phys. Rev. D  52, 186 (1995); arXiv: hep-ph/9501281.
\bibitem{char} J. Beringer et al. (Particle Data Group), Phys. Rev. D 86, 010001 (2012).
\bibitem{longbir} M. Beneke, G. Buchalla, M. Neubert and C.T. Sachrajda, Eur. Phys. J. C 61, 439 (2009); arXiv: 0902.4446[hep-ph].
\bibitem{longiki} A. Khodjamirian, Th. Mannel, A. A. Pivovarov and Y. M. Wang; arXiv: 1006.4945[hep-ph].
 \bibitem{bir} I. I. Balitsky, V. M. Braun, and A. V. Kolesnichenko, Nucl. Phys. B 312, 509 (1989).
\bibitem{iki} V. M. Braun, arXiv: hep–ph/9801222 (1998).
 \bibitem{uc} I. I. Balitsky and V. M. Braun, Nucl. Phys. B 311, 541 (1989).
\bibitem{aliev} T. M. Aliev, K. Azizi, M. Savci, Phys. Rev. D 81, 056006 (2010); arXiv: 1001.0227[hep-ph].
\bibitem{col} P. Colangelo, A. Khodjamirian, ”At the Frontier of Particle Physics/Handbook of QCD“, edited by M. Shifman (World Scientific, Singapore, 2001), Vol.3, p. 1495.
\bibitem{btoxs} B. Aubert, et al., BaBar Collaboration, Phys. Rev. Lett. 93 081802, (2004).
\bibitem{btoxssec} M. Iwasaki, et al., Belle Collaboration, Phys. Rev. D 72 092005, (2005).
\bibitem{btokstar} J.P. Lee, et al., BaBar Collaboration, (2012); arXiv:1204.3933 [hep-ph],.
\bibitem{btokstarsec} R. Aaij, et al., LHCb Collaboration, Phys. Rev. Lett. 108, 181806 (2012); arXiv:1112.3515 [hep-ph].


\bibitem{helicity} T. M. Aliev, M. Savc{\i}, JHEP 0605, 001 (2006); arXiv: hep-ph/0507324.
\bibitem{ampform} P. Bialas, J. G. K\"{o}rner, M. Kr\"{o}mer and K. Zalewski, Z. Phys. C 57, 115 (1993).
\bibitem{polden} J. G. K\"{o}rner and M. Kr\"{o}mer, Phys. Lett. B 275, 495 (1992).
\bibitem{2} T. Mannel and G. A. Schuler, Phys. Rev. D 279, 194 (1992).
\bibitem{3} M. Tanaka, Phys. Rev. D 47, 4969 (1993).
\bibitem{4}  M. Gremm, G. Koepp and L. M. Sehgal, Phys. Rev. D 52, 1588 (1995); arXiv: hep-ph/9502207.
\bibitem{5} C. S. Huang and H. G. Yan, Phys. Rev. D 56, 5981 (1997).
\bibitem{6} J. G. K\"{o}rner and D. Pirjol, Phys. Lett. B 334, 399 (1994); arXiv: hep-ph/9405360.
\bibitem{AlievSirvanli} T. M. Aliev, M. Savci, B. B. Sirvanli, Eur. Phys. J. C 52, 375 (2007), arXiv: hep-ph/0608143.
\bibitem{alievbashiry} T. M. Aliev, V. Bashiry, M. Savci, Eur. Phys. J. C 38, 283 (2004), arXiv: hep-ph/0409275.
\bibitem{Bensalem} W. Bensalem, D. London, N. Sinha and R. Sinha, Phys. Rev. D 67, 034007 (2003), arXiv: hep-ph/0209228.


\end{thebibliography}
\end{document}